\documentclass[envcountsame,runningheads,notitlepage]{llncs}

\usepackage{longtable}

\usepackage{adjustbox}
\usepackage{fancyhdr}
\usepackage{hyperref}
\usepackage{listings}
\usepackage{float}
\usepackage{pgf-umlsd}
\usepackage{tikz}
\usepackage{mdframed}
\usepackage{multirow}
\usepackage{array}
\usepackage{placeins}
\usepackage{array}
\usepackage{amsmath}
\usepackage{multicol}
\usepackage{msc}
\usepackage{tabularx}
\usepackage{booktabs}

\usepackage[utf8]{inputenc} 
\usepackage{graphicx} 

\usepackage{booktabs} 
\usepackage{array} 
\usepackage{verbatim} 
\usepackage{subfig} 

\begin{document}

\title{Post Quantum Migration of Tor}

\author{%
Denis Berger\inst{1}  \and 
Mouad Lemoudden\inst{1} \and
William J Buchanan\inst{1}
}

\institute{
Blockpass ID Lab, Edinburgh Napier University, UK,\\
\email{b.buchanan@napier.ac.uk}}


\maketitle




\begin{abstract}
Shor's \cite{365700} and Grover's \cite{grover1996fastquantummechanicalalgorithm} algorithms' efficiency and the advancement of quantum computers imply that the cryptography used until now to protect one's privacy is potentially vulnerable to retrospective decryption, also known as \emph{harvest now, decrypt later} attack in the near future. This dissertation proposes an overview of the cryptographic schemes used by Tor, highlighting the non-quantum-resistant ones and introducing theoretical performance assessment methods of a local Tor network. The measurement is divided into three phases. We will start with benchmarking a local Tor network simulation on constrained devices to isolate the time taken by classical cryptography processes. Secondly, the analysis incorporates existing benchmarks of quantum-secure algorithms and compares these performances on the devices. Lastly, the estimation of overhead is calculated by replacing the measured times of traditional cryptography with the times recorded for Post Quantum Cryptography (PQC) execution within the specified Tor environment. By focusing on the replaceable cryptographic components, using theoretical estimations, and leveraging existing benchmarks, valuable insights into the potential impact of PQC can be obtained without needing to implement it fully. 
\end{abstract}


\section{Introduction}
\subsection{Background}
To highlight the significance of the cryptography within the Tor network, a brief history and explanation of its functioning are needed. Released in 2002, Tor represents about two million estimated users \cite{TorMetrics}. It aims to fight censorship and surveillance, allowing users to browse anonymously. The definition of Tor's anonymity is described by Dingledine
and Syverson \cite{2GEN} as unlink-ability between the user's identity and their actions. Their paper highlights the "not perfect" side of the definition, explaining that it aims to prevent traffic analysis to a certain degree. Tor's mode of operation mainly relies on three types of "proxies", through which the user connection goes. The client selects an entry node, middle node and exit node, forming a path to the targeted resource. For each node, a key exchange happens between the client and the node, starting with the entry relay and then extending to the other nodes. Once the key exchange has occurred, data is re-encrypted at each step of the circuit. It makes three layers of encryption until it reaches the exit node, where the traffic is decapsulated to reach its final destination. 

The Tor network is often associated with facilitating illegal activities, primarily through \emph{onion services} - formerly known as hidden services. However, metrics collected by the Tor paper reveal a discrepancy between the proportion of onion service traffic and that of the total network bandwidth. While unlawful content is hosted on Onion services, this constitutes only a small fraction of the overall activity enabled by the Tor network (see Section \ref{tormetrics}). \hyperlink{OS}{OS} are a way to present content within the network itself without having to leave Tor. They conceal the physical location of the server and are accessible exclusively through the network using addresses ending in '.onion'. Since the implementation of Version 3, these 56-character alphanumeric strings are created by encoding the service's long-term public key (32 bytes Ed25519), checksum and version ("x03" by default) in Base32 \cite{torpropo}\#224. Dingledine
and Syverson \cite{2GEN} introduced a new mechanism called rendezvous-point relying on a DH key exchange where the user and OS establish a shared secret through Diffie-Hellman, with a "rendezvous" relay connecting the circuits from both parties without learning their identities or reading the data transmitted. Yet, since version $0.3.2.1-alpha$, Onion Services cryptography has been improved by replacing the schemes SHA1, DH, 1024-RSA with SHA3, Ed25519, curve25519. Today the main usage of cryptography in the network as shown in Table \ref{tab:relayKeys}, \ref{tab:OnionServiceKeys} is relying on RSA, ECC and AES. The latter, as a symmetric encryption algorithm, is less vulnerable to QC. An attacker equipped with a sufficiently advanced quantum computer could, in theory, compromise these cryptographic schemes to impersonate relay nodes, decrypt confidential content, and forge digital signatures, thereby creating documents that appear authentic to recipients. Gidney and Ekerå  \cite{Gidney_2021} estimate 20 million required physical qubits to break RSA-2048 encryption and highlight a considerable decrease in this estimation from 2015 to 2021, thus encouraging a faster adoption of PQC. 

\subsection{Purpose}

On the 13th of August 2024, the NIST \cite{FIPS204} released three long-awaited PQC standards, of which CRYSTALS-Kyber KEM and CRYSTALS-Dilithium signature scheme (renamed "ML-KEM" and "ML-DSA") both based on lattice problems. Another lattice-based standard publication, \hyperlink{FALCON}{FALCON} is expected shortly. On the other hand, the NIST more recently announced a new timeline, banning RSA and ECDSA from 2035 \cite{nist-ir-8547}. Along the wave of new publications in the field, the interrogation below persists:\\ \\ How can one evaluate its relay performance now as a reference point for later? \\ \\
The four core aims of this paper can be summarised in the following points to assist in answering the latter research question.

\begin{itemize}
    \item First, based on the literature analysis, the paper intends to identify a set of appropriate schemes for Tor post-quantum cryptography migration.
	\item To set up a functioning Tor network comprised of multiple physical nodes in order to measure it and evaluate its performance at scale.    
    \item Thirdly, identifying and measuring the cryptographic processes within Tor software.
    \item Finally, the scope of this research is to estimate a theoretical overhead value by incorporating the results of the previously defined objectives with post-quantum benchmark performances made on the devices used.
\end{itemize}

\section{Literature Review}

Tor specifications are supplied by proposals, allowing its implementation to grow and adapt to attacks over time. This literature review focuses on the cryptography features of the Tor paper. Firstly, the network's cryptography design must be explored to understand the technical environment and constraints. The second section of this review will go through the main schemes used, reviewing the relevant cryptosystems and their place within onion routing in order to identify the ones that could be broken by a cryptographically relevant \hyperlink{QC}{QC} and the potential replacement candidates.

\subsection{Tor Design and Architecture}
In 2004, Dingledine and Syverson \cite{2GEN} addressed the disadvantages of the initial design by introducing features that have remained fundamental to this day. One of them is directory authorities: they are a small group of "trusted nodes" reaching a common agreement on the network state.
In fact, Tor is not a fully decentralised, peer-to-peer network. This consensus allows Tor clients to have a global view of the network and aims to prevent malicious relays from manipulating its topology. Dingledine and Syverson \cite{2GEN} explain how these servers provide this information by sharing a directory signed with their long-term keys. Yet, the paper's technical design has been improved many times since. For instance, shortly after, the \hyperlink{TAP}{TAP} was introduced \cite{PET2006} and strengthened the cryptographic rigour formalising a circuit-based approach.

\subsubsection{Circuits and Handshake}
Circuit creation relies on the directory authorities in the very first step of the process when a given \emph{client} (OP) chooses the first relay. As it needs a list of reliable relays, it will fetch the consensus document from the directory authorities, providing necessary data on the relays, such as their public keys, roles, supported protocols and bandwidth \cite{torspec}. Figure \ref{fig:circuitcreation} below shows a high-level view of the circuit creation process. A circuit is typically composed of four channels: OP to R1, R1 to R2, R2 to exit node (RN) and RN to destination. After a circuit is established, the OP securely shares a unique symmetric key with each relay in the circuit. Additionally, each R exchanges a distinct circuit identifier with the nodes adjacent to it within the circuit, facilitating secure and efficient communication between nodes without revealing the entire path. A channel is defined as an encrypted link established directly between two relays or between a client and a relay, implemented as TLS sessions over TCP.

\begin{figure}
    \centering
    \includegraphics[width=1\linewidth]{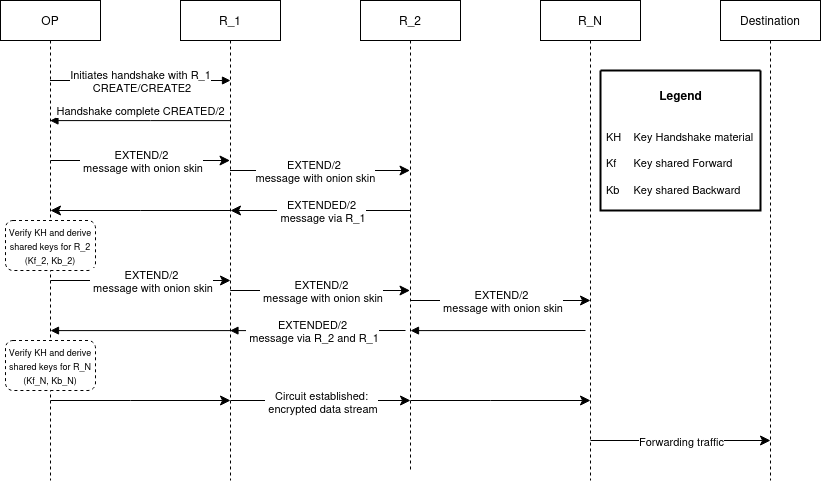}
    \centering
    \caption{Tor Circuit Creation. \textit{The /2 represents the version, as their older specifications became obsolete \cite{torspec}.}}
    \label{fig:circuitcreation}
\end{figure}

Tor relies on several cell types: Cells, shown in Figure \ref{fig:circuitcreation}, are responsible for circuit creation. During channel negotiation, CERTS cells are used to describe the keys that a Tor instance is claiming to have. It also provides certificates to authenticate that those keys belong to long-term key(s) that uniquely identify a relay.

In standard \hyperlink{TLS}{TLS}, the authentication is usually only necessary on the server side as the client identity is not important; this also applies to the first connection of a Tor circuit. Yet, the three other channel establishments need mutual authentication as the nodes must prove the genuineness of their affiliation in the circuit \cite{circuitExtHandshake2015}.

NTOR, the key exchange protocol currently used by Tor, introduced significant improvements over the TAP protocol \cite{NTOR}. It reduces the computational overhead by employing \hyperlink{ECC}{ECC}, offering faster key exchange and smaller data transmissions. Also, NTOR strengthens \hyperlink{FS}{FS}, preventing decryption, even if long-term keys are compromised. The paper reflects on the challenge of One-Way Authenticated Key Exchange (1W-AKE) and how it applies to anonymity. In 1W-AKE, one party (the client) authenticates the other(relay) and remains unauthenticated.  Goldberg et al.'s design \cite{NTOR} offers to authenticate the relay to the client by using the relay's public key and performing a key exchange via Curve25519 Diffie-Hellman.

Gosh and Kate \cite{cryptoeprint:2015/008} address the limitations of NTOR in light of quantum algorithms while maintaining forward secrecy, calling their 1W-AKE implementation HybridOR. It combines lattice-based cryptography, Learning With Error (ring-LWE), with the current DH assumption, maintaining compatibility with the current infrastructure.

Schanck et al. \cite{circuitExtHandshake2015} proposed a hybrid design of the ntor circuit-extension handshake demonstrating a practical implementation using \hyperlink{NTRU}{NTRU}Encrypt.
Schanck et al. \cite{circuitExtHandshake2015} permits the incorporation of any number of KEMs, allowing the other quantum-resistant algorithms to be integrated more easily without rebuilding the entire protocol. They compare the performance of this hybrid handshake with the main Tor's handshakes (tap and ntor) and the one from Gosh and Kate. Computation time (in microseconds, \(\mu s\)) and the communication overhead, the number of bytes transmitted between the client and server are shown for an instantiation with ntruees443ep1. NTOR remains the most efficient both in overhead and time taken. They also display the proportion of the total handshake time that is spent on the client-side operations relative to the overall handshake time. ntor equally shares the handshake time, the hybrid reaches 74\% and Gosh-Kate 67\%.

Gosh-Kate protocol's implementation by Schanck et al. \cite{circuitExtHandshake2015} showed, on average, a smaller computation time (900\(\mu s\)) but a larger number of bytes (1344). On the other hand, their hybrid design heavily relies on the client role (661\(\mu s\) for client init and 74\%), which could increase latency if the client's device has limited resources. However, their hybrid design implementation reduces the number of computational steps and amount of data exchanged. The evaluation is limited to NTRUEncrypt only; comparing it with other quantum-robust schemes would shed light on the efficiency of the overall hybrid implementation. They highlight a major obstacle, the size of the circuit-extension handshakes "CREATE" cells being limited to 505 bytes (in NTOR) while ntruees443ep1 requires 693 bytes \cite{circuitExtHandshake2015}. The work resulted in two distinct specification proposals; one aiming to widen the cell size \cite{torpropo}\#249, the other to enable the hybridisation of ntor protocol and a KEM. The first one has been supersuded by proposal \#340, introducing the sub-protocol "RelayCell" which focuses on cell packing and fragmentation. The second, \cite{torpropo}\#263, was made obsolete by proposal \#269; created by the authors of the papers and Tor's developers, it takes the incorporation of different post-quantum KEM further and emphasises compatibility with Tor’s handshake.

Currently, the Tor code includes four different circuit-extension handshakes \cite{torspec}:
\begin{itemize}\itemsep0pt
\item "CreateFast" deprecated (unauthenticated, non-forward-secure) handshake, which was previously used for the first hop of each circuit.
\item The ntor handshake. \cite{NTOR}
\item The Onion Service ntor handshake variant allows each party to encrypt data (without forward secrecy) after the first message. Clients have used it since version 3 of the OS protocol to encrypt data in the introduction and rendezvous cells. \cite{torpropo}\#224
\item The ntor v3 also permits each party to encrypt data at the cost of FS, enables the client to send an authenticated encrypted message within its onion skin and allows the relay to send an encrypted and authenticated reply as part of its response. \cite{torpropo}\#332

In 2019, Lauer et al. \cite{Lauer2020T0RTT} proposed a 0-\hyperlink{RTT}{RTT} handshake relying on puncturable KEMs to achieve a lower latency design than ntor; yet, the paper shows that it can result in higher computational overhead on certain low-power devices. The authors claim to achieve "immediate" FS, which appears to meet Tor's need for perfect FS, as the immediate variant mandates the deletion of ephemeral keys instantaneously. This approach could facilitate key management and allow the integration of post-quantum KEMs within the Tor handshake.
\end{itemize}

\subsubsection{Relay and Link Layer} 
\label{relaylink}
Also called onion skinning, the Relay "layer" is one of the most important of Tor's protocols. At the origin OP, a symmetric key is shared with each relay in the circuit using a telescoping key exchange protocol. During the circuit creation \ref{fig:circuitcreation}, instead of encrypting an entire circuit in one go, the circuit is built incrementally, with the client negotiating session keys with one node at a time. The OP encrypts the message multiple times, starting with the key for the final relay and moving backwards. When a message is transmitted, each intermediate relay decrypts one layer using its symmetric key, revealing the next hop or, for R\_N, the destination. The message is always padded to a fixed size, preventing traffic analysis based on message length. This approach ensures that no single relay has a full view of the communication path, enhancing anonymity. However, according to Degabriele
and Stam \cite{Degabriele2018}, due to its use of AES-128 in Counter-Mode for each layer of encryption, Tor's relay protocol is susceptible to tagging attacks, where a malicious entry and exit node can tamper with and detect changes in the data to de-anonymise the user. This is exacerbated by Tor’s low-latency design, which prioritises performance over stronger cryptographic guarantees.

Rogaway and Zhang \cite{OnionAE} introduce the concept of Onion Authenticated-Encryption, including indistinguishability from random bits and end-to-end authenticity verification as critical security measures. 

While Degabriele
and Stam \cite{Degabriele2018} assess the relay protocol under \hyperlink{CCA}{CCA}, identifying potential metadata leakage or improper handling of intermediary nodes, Rogaway and Zhang's findings \cite{OnionAE} highlight that Tor’s current protocol fails to meet these security benchmarks, specifically showing vulnerability to tagging attacks due to its reliance on counter-mode AES with an absence of a mechanism for authenticity checking throughout the relay path. They suggest improving integrity validation by applying robust authenticated encryption with associated data (AEAD) schemes which would ensure the detection of tampering at an intermediary (middle node) layer. Moreover, they propose to add layer-specific nonce usage and key diversification across encryption layers against tagging attacks. Globally, their work calls for a protocol restructuring, which would not be solved by the implementation of PQC. Since Grover's algorithm \cite{grover1996fastquantummechanicalalgorithm} demonstrates limited parallelisation capabilities, the quantum threat is not estimated to be significant to asymmetric cryptosystems; thus, increasing the key size is expected to provide sufficient security \cite{jaques2020grover}.

Concerning the link layer, TLS ensures authentication, integrity, and encryption of data in transit between nodes. Its speed is a crucial metric in the Tor protocol for the user experience as it directly impacts circuit creation latency (every new Tor circuit requires a fresh TLS handshake between nodes) and the network scalability as a higher TLS connection capacity allows to handle more users efficiently. At present, Tor's TLS avoids session resumption for additional security and relies on stateless connections to avoid potential state-carrying \cite{torspec}. Hence, treating each connection independently makes it harder for one to track connections or identify patterns. The newer version omits client certificates and uses a single-element, non-distinctive certificate chain to avoid detection by deep packet inspection systems, mimicking HTTPS traffic. TLS renegotiation was added but later moved to encrypted data records in version 3 to improve anti-blocking features and limit observable TLS signatures.

Tor aims to cease using TLS 1.2 \cite{torpropo} \#294, as 1.3 incorporates significant improvements such as the handshake design, particularly with its 1-\hyperlink{RTT}{RTT} meaning only one round-trip time is required until the first application message is sent, decreasing latency \cite{Dowling2021}. In a "classical" key exchange context, the assumption relies on Diffie-Hellman, while with PQC integration, it typically includes KEM, implicating a revision of the 1-RTT mode. In 2020, Schwabe et al. \cite{PQCTLSwithouthandshakeSignature2020} introduced "KEMTLS", a fully post-quantum modification of TLS 1.3, replacing signature-based authentication with KEM. Their design drastically cuts the computational costs associated with PQC signatures while achieving IND-CCA but not strict full-forward secrecy, for which they declared achieving levels 1, 3 and 5 according to the Noise protocol framework \cite{perrin2018noise}. The same authors released a variant of their previous KEMTLS named KEMTLS-\hyperlink{PDK}{PDK} \cite{cryptoeprint:2021/779}, yet in the context of Tor, this revised version relying on pre-distributed keys and partially cached information, might not be a straightforward fit if connections could be linked back to cached or pre-shared keys.

As for hybrid instantiation, the main method consists of concatenating the classical and PQC key material (public keys/ciphertexts) and treating them as a single element. Stebila et al. \cite{stebila2024hybridtls} draw a transitional capable design with backwards compatibility, allowing one device (client or server) to still use traditional schemes if it is not "hybrid-aware", therefore resulting in three possible scenarios: hybrid handshake, client downgrade or server downgrade to classical only. Their design aims to keep TLS 1.3 features such as high performance or 1-RTT.

The IETF draft declares the main security property of KEMs as IND-CCA2, correlating with the first motivation of CECPQ2 \cite{langley2018cecpq2}; the second version of the TLS 1.3 key exchange protocol developed by Google and Cloudflare. In fact, Langley \cite{langley2018cecpq2} highlights that, in TLS, managing confidentiality is more straightforward than authenticity, as encryption keys are independently negotiated for each session, while post-quantum authenticity presents a greater challenge, as it needs to integrate with the existing certificate authority and certificate ecosystem, making it considerably more complex to establish and maintain. 
Tor's TLS connection layer for relays and bridges, like the standard TLS, uses X.509 certificates to authenticate themselves during handshake exchanges \cite{torspec}. Concerning the OS, the ".onion" address, being a hash of the service's public key, confirms the server's identity itself without the need for a \hyperlink{CA}{CA}. Firefox, on which Tor's browser is based, has already released the option to use X25519Kyber768 for TLS.
Tor's TLS implementations have been slightly modified from the standard to meet its needs for compatibility and against traffic analysis. The ongoing research and deployments give a strong base to migrate the network's link layer; if the circuit-layer protocol ensures the main security and anonymity guarantees, in Tor, TLS is a necessary complement and has mitigated critical bugs in the past \cite{torpropo}\#294. The relay and circuit-extend protocols, as Tor-specific protocols, seem to need a deeper analysis to build a migration design. However, the relay layer does not appear highly endangered \cite{Baseri2024}. The stream protocol has not been reviewed due to its significant reliance on the other layers and its minimal impact on onion routing compared to them.

\subsection{Schemes}

Based on research papers and Tor's official source code, this section outlines the used schemes and their potential replacement candidates, each addressing specific security requirements. NIST defined PQC security level from 1 to 5. The first level is equivalent to the 128-bit traditional security level (strength of AES-128). 

TAP \cite{PET2006} implemented at most 80-bit security standard, \cite{nist80057}, through the use of RSA with a 1024-bit modulus and DH 1024 in its key exchange. Since NTOR \cite{NTOR} paper, Tor cryptography publications \cite{cryptoeprint:2015/008,circuitExtHandshake2015} agree on the necessity of at least a 128-bit security standard. Tor has started to adopt Curve25519 (also written x25519) in most key-enabled cryptography processes; nonetheless, RSA is still used as an identity key in the relay layer, see Table \ref{tab:relayKeys}, and in TLS 1.2.

Baseri et al. \cite{Baseri2024} compiled a comparative inventory of the vulnerable protocols and evaluated the corresponding mitigation strategies. The paper presents a risk assessment framework that includes multiple known attack strategies based on the STRIDE (Spoofing, Tampering, Repudiation, Information Disclosure, Denial of Service, and Elevation of Privilege) model. They differentiate the risk into two categories: the algorithmic level, where a mathematical assumption is not strong enough, and the protocol level, where the implementation presents a vulnerability, such as side-channel or fault-injection attacks. Hybrid implementation, aiming to protect the new cryptography migration from these risks, is highlighted in the context of a crypto-agility approach. However, due to Tor's infrastructure, the computational load, packet size, and bandwidth requirements need to be considered.

Table \ref{tab:tableschemes} breaks up Tor's layers by schemes (currently used), their security level, key size in bytes and role. When looking at the replacements, the benchmarking of PQC schemes tends to highlight the overall performance of lattice-based algorithms: Saber, Kyber as KEM and Dilithium as a signature scheme in constrained environments, such as running on Raspberry Pi \cite{pqcanalysisTLS}. Nevertheless, Saber was judged less efficient than Kyber in a broader range of environments \cite{Alagic2022}.

\begin{table}
    \centering
    \caption{Tor's main layers cryptography schemes.  \textit{*The "directory authorities" section is presented within the relay layer for clarity; however, it may be more appropriately classified as a distinct category. For further specifications, refer to Appendix \ref{Scheme specification}}}
    \label{tab:tableschemes}   
    
    \includegraphics[width=1\linewidth]{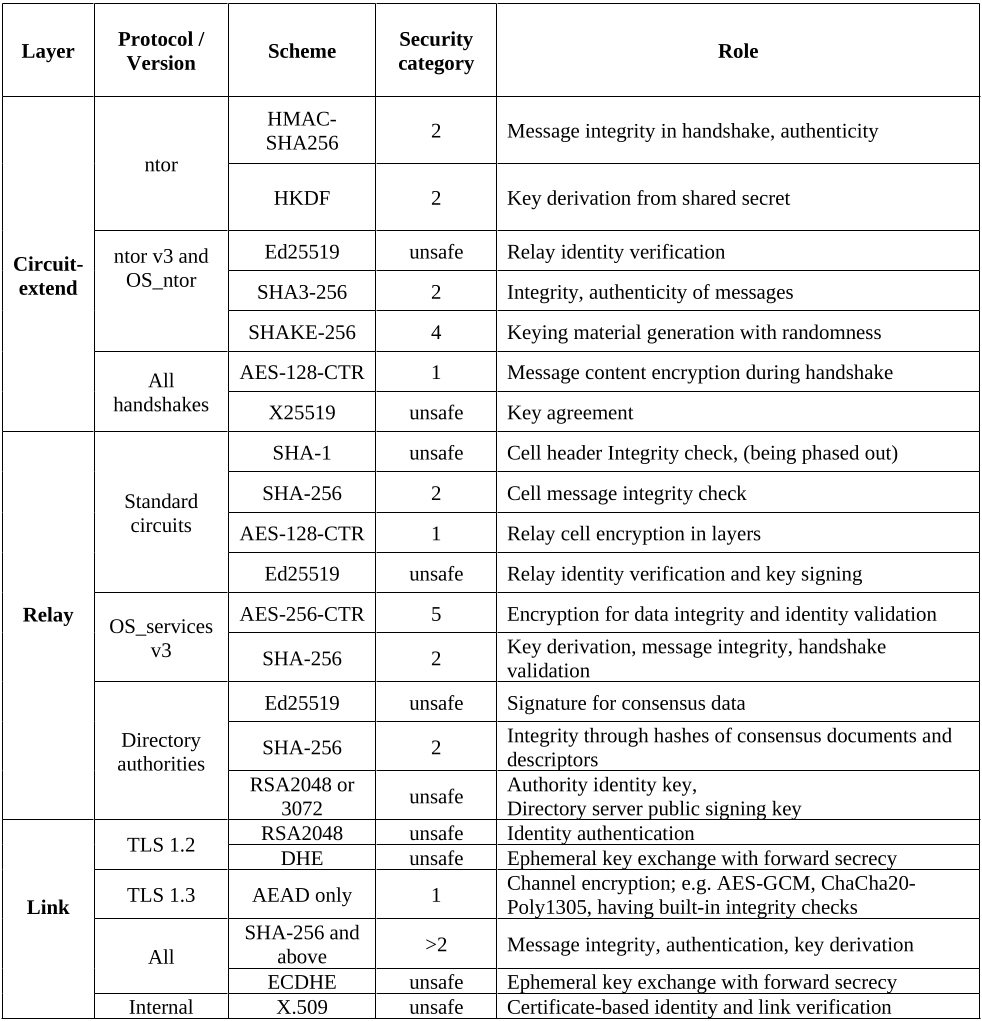}
 
\end{table}

The following first covers quantum-safe key exchange schemes, followed by a review of PQC signature algorithms.

\subsubsection{Key Exchange}
\label{KElit}
For key exchange, Tor almost always uses X25519 (Tables \ref{tab:tableschemes} and \ref{tab:relayKeys}). It is known for its efficiency and is used in multiple PQC hybrid instantiations \cite{stebila2024hybridtls,Kret2023pqxdh}.  The work of Schanck et al. \cite{circuitExtHandshake2015} is also a hybridisation of X25519 with ntruees443ep1 and HKDF-SHA256. In proposal \#269 \cite{torpropo}, a hybrid handshake alternative, it mainly differs from ntor in the computation of the authentication tag and key derivation.

Two lattice-based schemes have been instantiated as examples in the proposal: 
\begin{itemize}
    \item NTRUE KEM with EESS443EP2 specific parameter set, estimated at 128-bit security for both traditional and quantum-resistant settings. The design declares the maximum message size $m =$ 49 bytes, the KEM public key length $|PK|$ = 615 bytes, and a KEM ciphertext size $|CT|$ = 610 bytes.
    
    \item NewHope KEM is declared with $|PK|$ = 1824 and $|CT|$ = 2048.
\end{itemize}

The chosen NTRU parameter EESS443EP2 is interesting as it proposes relatively small keys; Cheng et al. \cite{Cheng2021} showed the performance of the \hyperlink{EES}{EES} parameters in constrained devices. Yet, according to  Bernstein et al., \cite{cryptoeprint:2016/461}, this classic NTRU parameter falls under a category vulnerable to automorphism-based exploits and other attacks such as lattice-reduction. In contrast, NTRU Prime is still of interest from a security point of view, achieving IND-CCA2; it is a third-round NIST candidate with its "Streamlined" and "LPRime" variants. The preferred parameter is Streamlined, sntrup761, as a balance of performance and security; it approximately achieves NIST category two post-quantum security with an estimation of $2^{153}$ based on the Core-\hyperlink{SVP}{SVP}. 

Another KEM scheme that was not selected for standardisation by the NIST in 2022 is FrodoKEM. For randomness generation, its main implementations rely either on AES-CTR or SHAKE XOF, the latter offers better performance across a broader range of hardware types, as demonstrated by Bos et al. \cite{Frodo2}. FrodoKEMs claimed categories 1, 3 and 5 of NIST PQC security according to the given parameters; FrodoKEM-640 targets the first level and FrodoKEM-976 the third. In Table \ref{fig:NTRUFrodosizes}, their key and ciphertext sizes are significantly larger compared to the sntrup761 setting except for the shared secret key. Furthermore, ML-KEM presents multiple advantages. It is indistinguishable from the chosen ciphertext attack, assuming that D-MLWE is intractable and that $G$, $H$, and $J$ are random functions. Additionally, it maintains IND-CCA2 by a quantum adversary able to make both classical and quantum queries (in superposition) to $G$, $H$, and $J$\cite{bos2018crystals,FIPS203}. Above all, it shows encouraging performances in comparison with other schemes \cite{pqcanalysisTLS,Kwala2024}. Table \ref{fig:MLKEMsizes} shows the three distinct standardised formats of Kyber and its corresponding key sizes. 

\begin{table}
    \centering
      \caption{Sizes (in bytes) of FrodoKEMs and sntrup761's keys and ciphertexts.}
    \label{fig:NTRUFrodosizes}
    \small
    \begin{tabularx}{\textwidth}{@{}l *{4}{>{\centering\arraybackslash}X}@{}}
        \toprule
        & \textbf{\scriptsize encapsulation key} & \textbf{\scriptsize decapsulation key} & \textbf{\scriptsize ciphertext} & \textbf{\scriptsize shared secret key} \\
        \midrule
        sntrup761  & 1,158  & 1,763 & 1,039  & 32 \\
        FrodoKEM-640-AES & 9,616  & 19,888 & 9,720  & 16 \\
        FrodoKEM-640-SHAKE & 9,616  & 19,888 & 9,720  & 16 \\
        FrodoKEM-976-AES  & 15,632  & 31,296 & 15,744  & 24 \\
        \bottomrule
    \end{tabularx}
 
\end{table}

\begin{table}
    \centering
        \caption{Sizes (in bytes) of keys and ciphertexts of ML-KEM.  \textit{\cite{FIPS203}}}
    \label{fig:MLKEMsizes}
    \small
    \begin{tabularx}{\textwidth}{@{}l *{4}{>{\centering\arraybackslash}X}@{}}
        \toprule
        & \textbf{\scriptsize encapsulation key} & \textbf{\scriptsize decapsulation key} & \textbf{\scriptsize ciphertext} & \textbf{\scriptsize shared secret key} \\
        \midrule
        ML-KEM-512  & 800  & 1,632 & 768  & 32 \\
        ML-KEM-768  & 1,184  & 2,400 & 1,088  & 32 \\
        ML-KEM-1024  & 1,568  & 3,168 & 1,568  & 32 \\
        \bottomrule
    \end{tabularx}

\end{table}

In hybrid implementations, the increase in handshake size tends to lead to higher bandwidth usage and latency during the handshake process. However, the overall performance heavily depends on the underlying mathematical assumption and its implementation. The Open Quantum Safe \cite{liboqs} benchmarking displays the performance of the schemes on x86\_64 architecture according to the operations (key generation, encapsulation, decapsulation) per second and per CPU cycle. It ranks HQC-128 and Kyber512, both achieving NIST level 1 security, with higher key generation per second. The table \ref{fig:OQSbench} shows a summary of OQS measures for the schemes of interest. Overall, taking into consideration size and performance attributes Kyber/ML-KEM appears to be the highest performer. It is followed by HQC, which reaches faster key generation per second and maintains good key encapsulation and decapsulation speeds despite larger key and ciphertext sizes (see Table \ref{fig:HQCsizes}). The streamlined NTRU Prime, sntrup761, comes next with smaller keys and ciphertext sizes than HQC-128, ML-KEM-768 or FrodoKEM-640. Claiming at least the second NIST security category, it represents a trade-off between performance, security, and bandwidth. FrodoKEM tends to fall behind due to reduced operational efficiency, considering its large key sizes and security level.

\begin{table}
    \centering
        \caption{KEM performance comparison. \textit{Operations per second per algorithm, Reference code type (unoptimised), 2024-04-02,  OQS \cite{liboqs}}}
    \label{fig:OQSbench}
    \small
    \begin{tabularx}{\textwidth}{@{}l *{5}{>{\centering\arraybackslash}X}@{}}
        \toprule
        & \textbf{\scriptsize keygen/s} & \textbf{\scriptsize keygen (cycles)} & \textbf{\scriptsize encaps/s}  & \textbf{\scriptsize decaps/s}  & \textbf{\scriptsize security category} \\
        \midrule
        HQC-128  & 6,719.33  & 371,892 & 3,571.00  & 2,349.00 & 1\\ 
        HQC-192 & 2,855.00  & 875,523 & 1,465.00  & 1,000.00 & 3\\ 
        Kyber512  & 23,348.00  & 106,974 & 18,970.00  & 16,118.00 & 1\\
        Kyber768  & 13,660.67  & 182,899 & 11,630.00  & 10,096.33 & 3\\
        sntrup761  & 115.55  & 21,636,049 & 2,308.00  & 818.00 & 2\\
        FrodoKEM-640-SHAKE & 345.99  & 7,224,951 & 282.67  & 280.33 & 1\\
        \bottomrule
    \end{tabularx}
\end{table}

While Kyber is a derivation of the Kyber PKE algorithm using \hyperlink{FO}{FO} transform, \hyperlink{HQC}{HQC} uses a variant of the FO transform called \hyperlink{HHK}{HHK}, allowing HQC to achieve IND-CCA2 security \cite{HQC2024}. Among a few other minor modifications, the FIPS 203 standard of Kyber, ML-KEM specifies a variant of the FO transform for the encapsulation and decapsulation mechanism; in this work, the Kyber benchmarks are considered for evaluating ML-KEM performance.

The code-based HQC scheme displays excellent performances, yet its large keys and ciphertexts (table \ref{fig:HQCsizes}) may require further adaptation to integrate it within Tor protocols.

\begin{table}
    \centering
        \caption{Sizes (in bytes) of keys and ciphertexts of HQC. \textit{\cite{HQC2024}}}
    \label{fig:HQCsizes}
    \small
    \begin{tabularx}{\textwidth}{@{}l *{3}{>{\centering\arraybackslash}X}@{}}
        \toprule
        & \textbf{\scriptsize encapsulation key} & \textbf{\scriptsize ciphertext} & \textbf{\scriptsize shared secret key} \\
        \midrule
        HQC-128  & 2,249  & 1,632  & 2,305 \\
        HQC-192  & 4,522  & 2,400 & 4,586 \\
        HQC-256  & 7,245  & 3,168 & 7,317 \\
        \bottomrule
    \end{tabularx}
\end{table}

The 512-byte limit affects the amount of data that can be included in the handshake process. The currently open proposal \#340 \cite{torpropo} aims to implement a cell packing, optimising the cell usage and a fragmentation mechanism allowing larger cryptographic keys.
Based on proposal \#269 design, the work on PQC migration toward a hybrid-handshake \cite{cryptoeprint:2015/008,circuitExtHandshake2015} and the attributes of the observed schemes, ML-KEM-768 and sntrup761 appear to fit the circuit-extension requirements. HQC may be an alternative if larger keys can be implemented easily. This work ignores strong cryptosystems such as McEliece, which imposes substantial key size and, therefore, communication overhead.

The current cell size limit remains an obstacle to hybrid or full PQC migration for all layers. Concerning the relay layer, no key exchange has been processed, but the challenge also concerns signature schemes. Regarding the link layer, it produces key exchange within its TLS implementation. The \hyperlink{RFC}{RFC} \cite{rfc8446} standard defines the handshake with three core stages: server parameters, key exchange and authentication. In a PQC environment, the second phase would rely on KEMs. The third is reviewed in the next section. Currently, Tor's TLS uses P256 ECDHE (and still has legacy support for secp224r1) for handshakes and Ed25519 for server identity known as link keys \cite{torspec}. Indeed, ML-KEM-768 appears to be the selected parameter for the newly adopted designs \cite{stebila2024hybridtls}. Færøy \cite{ahf_torfork}, in a fork of C Tor, experimented the integration of hybrid key exchange TLS with \verb|X25519Kyber768Draft00| . 

In 2020, Paquin et al. \cite{TLS_PQC} presented the performance trade-offs of several schemes, including hybrid ECDH-P256-Kyber512. Compared to the classical p-256 curve, the hybrid scheme shows an increase of packet loss and completion time. Their work highlights the effect of higher RTTs, where the hybrid scheme demonstrates a greater degradation; in their conclusion, the authors discussed how the enlargement of \hyperlink{MTU}{MTU} might improve TLS establishment performance. In 2024, the post-quantum TLS survey by Alnahawi et al. \cite{CiC-1-2-6} points out the significant impact of large keys with FrodoKEM as an analogy with 2.7× the time of a Kyber handshake and 2.53× the time of the typical classical handshake with x25519. They state that the combination of Kyber512 (NIST security level 1) and x25519 took 1.25× the time of the traditional handshake, while the NIST level 3 of this hybridisation has a very close performance (1.28×). Finally, the paper overviews how the work of \cite{PQCTLSwithouthandshakeSignature2020} differs from other designs in its approach of achieving AKE as the pure PQC solution KEMTLS. Indeed, KEMTLS is achieving a form of 1W-AKE as the server is authenticated to the client without DSA by using a long-term KEM public key for encapsulation, with the server responding via an encapsulation based on the client's ephemeral KEM public key. 

Therefore, achieving a 1W-AKE means the client does not need to authenticate itself. A pure KEM 1W-AKE implementation is interesting in the context of the future circuit-extension handshake. Recently, Pan and al. \cite{cryptoeprint:2024/361} introduced a one-way Verifiable Weak FS notion and presented the first lattice-based tightly FS AKE via key confirmation in the classical random oracle model (using the lattice-based protocol from Pan et al. \cite{cryptoeprint:2023/1380}); showing that the OW-VwFS can be transformed $tightly$ to FS using key confirmation in the random oracle model (ROM). The circuit-extension handshake requiring FS \cite{circuitExtHandshake2015}, in a near future a fully KEM reliant 1-WAKE handshake could potentially replace ntor or the hybrid-ntor. Based on \cite{PQCTLSwithouthandshakeSignature2020}'s benchmarks (NIST Round 3 estimation results), it could be as fast as or faster than the current handshake encryption.

\subsubsection{Signature}
\label{SIGlit}
This subsection takes a closer look at the potential signature mechanisms that could be integrated within Tor's different layers. 

The current 1W-AKE relies on public key cryptography. The hybrid circuit-extension design presented in proposal \#269, inspired by Schanck et al.’s work \cite{circuitExtHandshake2015}, does not integrate post-quantum DSA. It maintains the usage of ECDH primitives (signing with the server's Curve25519 public key) for authentication. This is based on the assumption that the session negotiation itself is not vulnerable to quantum attacks. The hybrid protocol derives session keys and authentication tags from shared secrets of both the ECDH shares and KEM-derived secrets, allowing the verification of authenticity without an explicit digital signature during session establishment. This approach avoids the additional computational cost and larger communication footprints of the standardised DSAs and the significant modifications to the Tor protocol that their incorporation would lead to. Although the recently introduced ML-DSA and SLH-DSA \ref{tab:DSA} have larger signature sizes compared to the stateful hash-based schemes \hyperlink{XMSS}{XMSS} and \hypertarget{LMS}{LMS} \ref{tab:DSA2020}, they offer the advantage of reducing state management complexity and lower operational overhead; thus are better candidates in case they would be used in the handshake. The Falcon scheme, like ML-DSA, is based on Lattice; it proposes even smaller keys (Table \ref{tab:FalconDSA}) but considerably slower run times; Dilithium with security level 2 parameters generates key pairs approximately 178 times faster per second than Falcon512, which corresponds to security level 1 (Table \ref{fig:OQSbench2}). SLH-DSA is the only stateless hash-based standard so far, offering robust security without relying on new mathematical problems. However, it generates large signatures (from nearly 8 KB to 50 KB) and performs slower signing than its lattice-based competitors; SPHINCS+ at security Level 3 achieves about 14 signing operations per second as shown in Table \ref{fig:OQSbench2}.

\begin{table}
    \centering
       \caption{PQC performance comparison.  \textit{Operations per second per algorithm, Reference code type (unoptimised), 2024-04-02,  OQS \cite{liboqs}}}
    \label{fig:OQSbench2}
    \small
    \begin{tabularx}{\textwidth}{@{}l *{5}{>{\centering\arraybackslash}X}@{}}
        \toprule
        & \textbf{\scriptsize keygen/s} & \textbf{\scriptsize keypair (cycles)} & \textbf{\scriptsize sign/s}  & \textbf{\scriptsize verify/s}  & \textbf{\scriptsize security category} \\
        \midrule
        Falcon-512  & 53.26 & 46,938,518 & 176.55  & 17,246.00 & 1\\ 
        Falcon-1024 & 17.91 & 139,565,458 & 80.56  & 8,341.00 & 5\\ 
        Dilithium2  & 9,486.00 & 263,444 & 2,099.33  & 8,752.33 & 2\\
        Dilithium3  & 5,184.33 & 482,134 & 1,320.00  & 5,519.00 & 3\\
        SPHINCS+-SHA2-128f-s & 558.81 & 4,473,508 & 23.86 & 404.73 & 1\\
        SPHINCS+-SHA2-192f-s & 382.08  & 6,542,247 & 14.52  & 270.91 & 3\\
        \bottomrule
    \end{tabularx}

\end{table}

The relay layer is heavily reliant on the circuit-extension handshake in terms of authentication, yet it also depends on Tor's own certificate mechanism. Tor does not rely on standard CA-issued certificates for its core functioning, avoiding reliance on centralised CAs. Signing is done with Ed25519 keys; their format differs for certificates used by authorities to sign their identity key \cite{torspec}. $CERTS$ cells are at least 104 bytes when containing a single certificate of 96 bytes (32 of certified key and 64 of signature), its expiration date of 4 bytes plus 4 bytes representing the number, type and length of certificates in the cell. This number is approximate as an extension can be added to bundle the signing key along with the certificate, which adds up to at least 36 bytes. Therefore, 140 bytes would represent one certificate and its signing key, which fits into the current absolute maximum fixed size of 512 bytes. ML-DSA, the smallest recently standardised post-quantum equivalent, where only the public key and signature sizes represent a total of 3,732 bytes (Table \ref{tab:DSA}). 

In regards to Onion Services, which also rely on a public key for authentication (Table \ref{tab:OnionServiceKeys}), to prevent the linking of descriptors, they use a blinded version of the identity key that changes at regular intervals instead of using the identity key directly \cite{torspec}\#224. This allows to hide the original identity key while still authenticating without linking the real key directly. The feature is vulnerable to quantum attacks where a capable adversary could forge the OS signature and potentially redirect queries addressed to the service. Regarding its migration to PQC primitives, Eaton et al. \cite{Eaton2021} assessed four schemes, including Dilithium, which outperforms the others both in signing and verification but also reaches a close result of its unblinded counterpart.

TLS signatures can be divided in two categories, "online" signatures of messages in the handshake protocol and static ones of
certificates in the certificate chain; where the "statics" allow a greater signing time as the computation takes place in advance. Typically, the root certificate signs the intermediate CA certificate, which itself signs the leaf certificate. The latter is used to sign the transcript during the handshake; the other signatures are static. The certificate chain size gives practical insight into assessing the overhead. Kampanakis and Childs-Klein \cite{kampanakis2024impact} estimate the authentication data size of ML-DSA-44 and ML-DSA-6 of 14 KB and 19 KB, respectively (including the intermediate CA certificate). Their work on the increased latency caused by the certificates is based on the estimated sizes of the chain; the observed chains are built with RSA, not actual ML-DSA certificates. The paper states a ~32\% handshake time increase compared to 2.5~KB chain. 
In Tor TLS, the certified key type varies from Ed25519 to a hash (SHA256) of an X.509 certificate. Since the hash itself is not directly reversible, this part of TLS authentication is therefore less exposed to quantum attacks. Yet, the underlying X.509 certificate remains dependent on the public key primitive used. While standards drafts are proposing to include the recently standardised SLH-DSA \cite{ietf-lamps-x509-slhdsa-02}, its performance stays far behind the Dilithium and Falcon \cite{cryptoeprint:2020/071}. Kampanakis
and Kallitsis \cite{WithoutCertificates} proposed a backwards-compatible mechanism to omit the intermediate CA certificate, allowing lighter and faster PQC TLS Handshakes. 

Currently, an IETF draft is in progress to extend \hyperlink{ACME}{ACME} challenges to validate ".onion" domains through a Tor-compatible mechanism; The draft by Misell \cite{misell2024acmeonion} presents a new challenge type, "onion-csr-01", while still incorporating the "http-01", "dns-01" and "tls-alpn-01" challenges. Relying on the broader web PKI, aside from the potential privacy concerns for the onion service, stays sensitive to quantum attacks. 

If proposal \#340 is implemented as specified, the digest field, responsible for checking the cell integrity, should reach a size of 14 bytes, allowing the integration of larger hashes (currently SHA-1). Although hash functions are less directly threatened by quantum algorithms than \hyperlink{PKE}{PKE}. It has been shown that Grover's algorithm can be used to reduce the time required for preimage attacks on hash functions like SHA-1, SHA-2, and SHA-3; and also to minimise the time needed for hash inversion in \(O(\sqrt{n})\), halving the security of the function \cite{Preston2022}. 

Ultimately, the NIST \cite{nist.ir.8528} is still looking for a signature scheme and declared that some second-round candidates have undergone minimal or no formal cryptanalysis in published research. Thus, potential signature schemes may still reveal themselves in future research and deployment as a match for Tor's requirements. For instance, FAEST, which achieves faster key generation than ML-DSA, displays smaller key sizes and relies on the mature security of SHA3 and AES, yet its signatures are more than twice the size of the latter lattice-based scheme for their respective minimum security levels \cite{FAEST}.

\subsection{Reflection}
The attraction towards hybrid models is justified by the potential fallback to classical encryption in case a given quantum robust scheme fails. Even if they have been cautiously examined, new PQC schemes could potentially be broken by an adversary. As an analogy, \hyperlink{SIKE}{SIKE} reached the fourth round of the NIST competition before being proven insecure \cite{cryptoeprint:2022/975}. To not solely rely on relatively young solutions, hybrids appear to outweigh the faster, fully PQC (KEMTLS). As for the ease of instantiation, the concatenation
method of algorithms allows a simpler implementation of these hybrids.

The circuit layer comes up as the most urgent layer to migrate to PQC. Implementing a transitional hybrid scheme would require a KEM and a DSA mechanism. The second migration would concern the link protocol. In comparison, TLS has benefited from a greater number of deployments and research. While OS reuses these fundamental blocks, it incorporates additional mechanisms such as introduction, rendezvous points and hidden service descriptor encryption, which have separate cryptographic processes. Public key cryptography is used among both: key exchange for the introduction, rendezvous points and authentication for both points and descriptors. 

Regarding Tor's components that should be tested with quantum-robust KEM, four are standing out:
\begin{itemize}
        \item The circuit-extension handshake
        \item Onion skin
        \item Link layer TLS
        \item OS Introduction, Rendezvous points
\end{itemize}

As for the ones that can be evaluated using DSA, the following stand out:
\begin{itemize}
        \item Consensus document
        \item Relay Identity Authentication
        \item Link TLS handshake authentication
        \item Encrypted Descriptor
\end{itemize}

The subsection \ref{relaylink} highlighted the fact that Tor already faces cryptography challenges that will not be solved by the integration of quantum-resistant schemes. The latter implies a potential redesign of certain underlying protocols beforehand, as proposed by Degabriele and Stam \cite{Degabriele2018}. Overall, the review emphasized the various cryptographic layers and looked over their interactions. It appears technically feasible to transition to PQC standards incrementally, layer by layer. Regarding the necessity of migrating the encapsulated TLS layer to PQC when its outer layer is already quantum-resistant, the recommendation is affirmative, as it is preferable to ensure comprehensive quantum security.

On the other hand, Rahman et al. \cite{Rahman2024} proposed an integration of \hyperlink{QKD}{QKD} for symmetric key exchange without relying on trusted nodes. The design presents a quantum relay to facilitate the key exchange between client and nodes, yet this approach would require a specific quantum communication infrastructure, such as quantum repeaters, which are still in development and not widely available.

It can be observed that the diversity of KEM standards is currently limited compared to that of DSA (ML-DSA, SLH-DSA, XMSS and \hypertarget{LMS}{LMS}).


\section{Methodology and Design}

This section seeks to develop the implementation design based on insights gained from the literature review. The research takes an inductive approach, collecting data to build a theory regarding the impact of PQC migration in Tor. 

\subsection{Technical Requirements}

On a security level, the implemented protocols will need to achieve IND-CCA2, and at least the first NIST PQC security category. 
From a size perspective, none of the schemes observed fit in the current cell size. At the data-link layer, to reduce fragmentation, it is preferable for public key sizes to fit in the Ethernet MTU of 1500 bytes. Only ML-KEM-512,768 or sntrup761 for KEM and Falcon-512, ML-DSA-44 for signature are matching. Yet, as previously noted, Tor cell body lengths are currently limited to 509 bytes and will be adjusted to 493 bytes, except for "DATAGRAM" messages which aim to support UDP-over-tor Tor proposal \cite{torpropo} \#339. Fragmentation of PQC keys across several cells appears inevitable (Tables  \ref{fig:NTRUFrodosizes}, \ref{fig:MLKEMsizes}, \ref{fig:HQCsizes}, \ref{tab:DSA}, \ref{tab:DSA2020}, \ref{tab:FalconDSA}). Reassembly, verification, and potential retransmissions will be necessary. ML-KEM and Falcon encapsulation and public keys respectively fit in two Tor cells. In terms of actual runtime, ML-DSA significantly outperforms Falcon.

Regarding relays, memory ($\geq 1$ to 1.5~GB of RAM per node), disk storage ($\geq 200$ MB) and bandwidth minimal requirements ($\geq 16$ Mbps) as defined by Tor paper will need to be met in the hardware equipment used. 

\hyperlink{ARTI}{Arti} offers a safer and faster development than C Tor due to Rust's design. Generally, Rust language does not outperform C, yet it shows close performances \cite{Zhang2022}. Initiated in 2020, the Rust implementation takes a simpler approach compared to the older C version. At present, a performance comparison of the two implementations is difficult to make as Arti is not fully finished. For instance, relays cannot be run just yet \cite{torspec}. Thus, this work focuses on C language Tor implementation and cryptographic libraries.

\subsection{Network Architecture}
\label{sec:netwarch}

A star topology will be formed around a central switch device. For the sake of mimicking the key roles, the network needs to provide enough diversity; therefore, nine nodes will operate together. 
Two machines will act as directory authorities. Four nodes will serve as guard relays, two as exit relays, and one will be dedicated to functioning as a middle hop and OS. Given the size and configuration of the network, there will be no dedicated directory caches, and nodes will endorse multiple flags. For instance, the onion service descriptor "HSDir" flag.

\subsection{Local Benchmark}

The implementation will start by looking at the tools relevant to measuring the Tor network and running them within our scenario. A packet capture will permit the observation of the time of transmitted frames with the corresponding traffic type. 

Tor uses \texttt{sbws} (Simple Bandwidth Scanner) and OnionPerf to measure its performance. \cite{TorMetrics}.
\texttt{Sbws} usually relies on the download and upload of files through the circuits to measure the benchmark. This method can be mimicked manually using tools such as \texttt{wget} or \texttt{curl}. The implementation will look for a method to efficiently measure the circuit round-trip latencies and circuit build times. 

\subsection{Code Performance}
\label{codeperfo}

With regard to the three layers, the list of C files below represents the core Tor protocol operations; they will be assessed on the performance of their most computationally expensive functions.

\begin{itemize}
    \item \texttt{circuit\_establish\_circuit} function in \texttt{circuitbuild.c} handles the process of selecting paths and managing the handshake protocols between nodes to establish circuits.
    \item  \texttt{onion.c} is responsible for the creation, encoding, and parsing of cells (CREATE, CREATED, EXTEND, and EXTENDED). It invokes handshakes.
    \item \texttt{channel.c} implements the transmission, reception, and processing of cells across different connections.
    \item \texttt{relay.c} manages RELAY cells, including their encryption, decryption, forwarding, and processing for routing data.
\end{itemize}

The following code files will be evaluated to weigh the cryptography operations they represent.

\begin{itemize}
    \item \texttt{onion\_ntor.c} implements the ntor handshake.
    \item \texttt{onion\_ntor\_v3.c}, as specified in proposal 340 \cite{torpropo}, the cell fragmentation requires the enhanced handshake, ntor v3. 
    \item \texttt{relay\_crypto.c} relay cells payload encryption, decryption, integrity verification via cryptographic digests, and the initialisation of symmetric keys.
    \item \texttt{hs\_ntor.c}, closer to ntor v3 than ntor; it is the implementation of the handshake for onion services.    
    \item \texttt{tortls.c} is the main TLS implementation file in Tor.
    \item \texttt{channeltls.c}, is the only instantiation of channel abstraction. It handles the v3+ link handshake, certificate verification, and cell processing over OR connections.
\end{itemize}

In Tor's "crypto" code directory, the deprecated fast handshake file \texttt{onion\_fast.c} is ignored. A frequency analysis will be conducted to associate roles of nodes with the functions they execute the most.

\subsection{PQC}

The following algorithms will be considered viable candidates: ML-KEM-512 and Falcon-512 at security level 1, as well as sntrup761, ML-DSA-44 at security level 2 and ML-KEM-768 at level security 3.  The Open Quantum Safe C library \cite{liboqs} provides implementations of the considered schemes. The speed tests for signature and KEMs will be run on the different devices with their code. Regarding the performance evaluation of the classical scheme, it will use the OpenSSL \texttt{speed} tool.
Combinations of the selected schemes will be made to evaluate their timing with their security level. All of them will be run on at least two different devices. Advertised results summarised in sections \ref{KElit} and \ref{SIGlit} will be taken into account. 

\subsection{Reflection}

The implementation of this outline should be helpful in shedding light on the weight of post-quantum in Tor according to the assumption that if a given cryptographic process represents a proportion of Tor's total run-time, then a given latency increase will lead to a raise of at least of the total timing.

\section{Implementation}

Firstly the benchmark techniques will focus on the local network before targeting cryptographic timings (circuit-extend, onion service and TLS). Secondly, the measurements obtained will be compared with their potential PQC substitutes.

\subsection{System Setup}
The hardware used for this implementation consists of 9 Raspberry Pis used as onion relays, of which two model 4b with 8~GB RAM, seven model 5 (two with 8~GB and five with 4~GB RAM). As a client, a Thinkpad laptop X1 with i7-10850H CPU and 32GB RAM. They are connected using two-metre-long straight-through Ethernet cables and a Cisco Catalyst 2900 switch (Figure \ref{fig:hardware}). It should be noted that a Raspberry 4b with 4~GB RAM is added for stability when needed.

Each Raspberry runs Ubuntu server 24.04, and the Lenovo personal computer runs Kali Linux 2024.3. The switch runs \hyperlink{DHCP}{DHCP} to avoid access the Raspberries through monitors, but the IPs are configured statically, the switch's minimal configuration can be found in Listing \ref{lst:ciscoconf}. Each node is running Tor version 0.4.8.10.

The Nginx web server is used for the hidden service. It publishes a simple http page \ref{fig:OStest}. It is resolved and reached by launching the Tor browser using a separate Tor daemon (see Listing \ref{lst:TorBrowsercommand}). Tor software is designed to run within a large network with a great number of nodes; the implementation of a local testing network with a small number of nodes meets multiple constraints. To ensure anonymity, Tor code discards guards on certain conditions.  

\subsection{Local Network Benchmark}

Average bandwidth connectivity between devices was measured to be 93.8~Mb/s using the iPerf network performance measurement tool \cite{iperf_tool}. The network is configured as described in Section~\ref{sec:netwarch}; Chutney \cite{torspec}, a Python tool for testing Tor network locally, is used to generate torrc configuration files for each node (Listings \ref{lst:chutconf} and \ref{lst:chutfiles}).

The torrc files have been slightly adapted, but most of the settings remain identical to Chutney's templates. Listings \ref{lst:authtorrc} and \ref{lst:authtorrc2} display the used parameters for directory authorities and guard relay. For instance, due to the size of the network setting \texttt{PathsNeededToBuildCircuits} has been decreased to allow more flexible circuit building. The client configuration also points to the local directory authorities and contains the setting for testing networks, which adjusts default values, as described in "man tor" \ref{lst:clienttorrc}. The conflux setting has been disabled, and the nodes have been specified in the configuration to facilitate the onion. 

If accessing the internet through the local Tor nodes, from the browser's point of view, the timing displayed is heavily influenced by external variables such as broadband and thus is not relevant to evaluating the Tor implementation itself. However, considering the timing difference between the same test request made from Firefox to the internet directly and the Tor browser going through the local nodes, it can be inferred that passing through the established Tor channel only adds a few hundred milliseconds \ref{fig:timingcomparison}. Tor's traffic is mostly web \cite{TorMetrics}, measuring the time taken to visit an internet resource adds a relevant data point of the overhead introduced by the class. The laptop is configured to perform \hyperlink{NAT}{NAT} between its ethernet and WLAN interfaces, and its IP is set as the default gateway on all nodes. \hyperlink{DNS}{DNS} points to Google 8.8.8.8 address. 

\subsubsection{Pcap Analysis}
This configuration allows sniffing the first packet sent to the guard relay and the first packet leaving the exit node \ref{fig:circuitOnBrowser}. (to the duckduckgo.com domain without search content) For this circuit exiting by node ".23" the client sends its Tor encrypted packet to the entry node at 15:29:50.360220332. The exit node sends a DNS request at 15:29:50.414655443 and its decapsulated \texttt{ClientHello} message at 15:29:50.462253502 (see Figure \ref{fig:sniffing}). Therefore, giving timing of 102 milliseconds to exit the circuit, including the DNS requests (which are influenced by broadband) and 54 milliseconds without.

When trying a similar request, the measured time, with "DNS overhead", is 110 milliseconds and approximately 60 milliseconds without (see Figure \ref{fig:sniffing2}). The browser process is killed between each measure, and its cache stays empty. In this capture, entry and exit nodes are the same, but the circuit is different, as shown in Figure \ref{fig:circuitOnBrowser2}. For a third circuit and another web request, we get 4~ms from the first TCP sent to the DNS request made by exit and 57~ms if waiting for the \texttt{ClientHello}. This timing further stresses the need to automate the measurement process to derive metrics from a larger sample size. 

\subsubsection{Simple Bandwidth Scanner}

As for the \texttt{sbws} tool, it is currently developed for targeting the real Tor or a locally hosted (single-device) Tor simulation. Here, it has been built locally, and its source code file, "generals.py" has been slightly modified to force the generated Tor daemon to point at the local directory authorities. This method avoids including the processing of the browser. To use this tool, a new node has been added to act as a web server. It publishes the 1~GiB file accessible on port 443 with HTTPS (TLS is required on the target in this scenario to run \texttt{sbws}). The laptop has been configured as a DNS server, allowing the nodes to resolve the address "local.tor" to IP 192.168.1.24. The self-signed \hyperlink{SSL}{SSL} certificate is updated as trusted on the nodes and the option \texttt{ServerDNSDetectHijacking 0} is added to their torrc configuration. The tool stops after one loop when detecting that it is assessing a testing network and creates a raw text file, which is used to generate "v3bw" files. As for OS, "Attachstream" functionality is not currently supported for .onion addresses. 

Gauging the circuit's timings against the direct and non-encapsulated traffic would bring a relevant comparison. Thus, the following subsection aims to do so.

\subsubsection{Curl Measurements}

First, we compare the latency of passing through the network with sending curl requests directly to the external server. The listings  \ref{lst:curl} and \ref{lst:curl2}, display the curl requests run from the client passing through its Tor daemon with SOCKS5.

\lstset{
  basicstyle=\scriptsize\ttfamily, 
  breaklines=true,                 
  frame=single,                    
  language=bash,                   
  showstringspaces=false
}
\begin{lstlisting}[caption=Curl GET request to duckduckgo,breakindent=0pt ,breaklines=true, label=lst:curl, float=hbt!]
    curl --socks5-hostname 127.0.0.1:9050 -w "%{time_total} " -o /dev/null -s https://duckduckgo.com
\end{lstlisting}

The Tor paper defines Round-trip latencies as the time between sending the HTTP request and receiving the HTTP response header \cite{TorMetrics}. Here, this is replicated by \texttt{time\_total} variable in the curl command. This technique also permits us to reach the OS and to directly benchmark it. 

\begin{lstlisting}[caption=Curl GET request to Onion Service, label=lst:curl2, float=hbt!]
    curl --socks5-hostname 127.0.0.1:9050 -w "%{time_total} " -o /dev/null http://adg4jkv2xpciraegkj3rpcfrdnzlaeqnlt7a3wzylhzcvshzkerhacyd.onion
\end{lstlisting}

\begin{figure}
    \centering
    \includegraphics[width=1\linewidth]{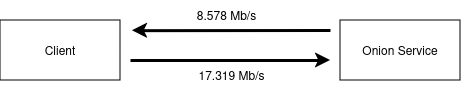}
    \caption{Onion Services download and upload average speeds}
    \label{fig:OSspeeds}
\end{figure}

Furthermore, POST requests are also sent to the target address, allowing us to download and upload performances with a 1~GiB large file (just as sbws can do) and the average of a typical POST request. This is done by adding curl parameters, as shown in Listing \ref{lst:curl3} for the evaluation of access to an external resource without relying on broadband internet. Figure \ref{fig:OSspeeds} represents the average speeds measured with the 1~GiB file. 

Finally, to evaluate the overhead of TLS in the onion service, it is implemented and compared with HTTP.

\subsubsection{Circuits Build Time}

Using \texttt{stem} library, the following script allows obtaining timing metrics for a given number of built circuits. Two ways have been implemented: the first is by waiting for the controller to mark the status of the circuit as built, and the second one is by checking its status at small intervals. The latter demonstrates greater consistency in its results.

\lstset{
  basicstyle=\scriptsize\ttfamily, 
  breaklines=true,                 
  frame=single,                    
  language=bash,                   
  showstringspaces=false
}
\begin{lstlisting}[caption=Python script using \texttt{stem} controller library, label=lst:stemscript, float=hbt!]
import time
import numpy as np
from statistics import mean, median, stdev
from stem import CircStatus
from stem.control import Controller

def benchmark_circuits(controller, numCircuits=1000):
    circuitBuildTimes = []   
    for i in range(numCircuits):
        startTime = time.time()
        # Creating a new circuit and waiting until circuit is fully built
        circId = controller.create_circuit(await_build=False)
        while True:
            circ = controller.get_circuit(circId)
            if circ.status == CircStatus.BUILT:
                break
            time.sleep(0.001) # Checking the status as frequently as possible
        endTime = time.time()
        buildTime = endTime - startTime
        circuitBuildTimes.append(buildTime)
        controller.close_circuit(circId)
        
   if circuitBuildTimes:
        avgTime = mean(circuitBuildTimes)
        mdTime = median(circuitBuildTimes)
        mnTime = min(circuitBuildTimes)
        mxTime = max(circuitBuildTimes)
        sdTime = stdev(circuitBuildTimes) if len(circuitBuildTimes) > 1 else 0
        q1Time = np.percentile(circuitBuildTimes, 25)
        q3Time = np.percentile(circuitBuildTimes, 75)
        print(f"Metrics for {numCircuits} circuits:")
        print(f"  Average build time: {avgTime:.4f} seconds")
        print(f"  Median build time: {mdTime:.4f} seconds")
        print(f"  Min build time: {mnTime:.4f} seconds")
        print(f"  Max build time: {mxTime:.4f} seconds")
        print(f"  Standard dev: {sdTime:.4f} seconds")
        print(f"  First quartile (Q1): {q1Time:.4f} seconds")
        print(f"  Third quartile (Q3): {q3Time:.4f} seconds")
    else:
        print("No circuits")

if __name__ == "__main__":
    with Controller.from_port(port=9051) as controller:
        benchmark_circuits(controller, numCircuits=1000)  
\end{lstlisting}

\subsection{Cryptography Benchmark}

To measure the cryptographic operations performance in terms of execution time, the relevant functions of the files described in section \ref{codeperfo} are executed with the \texttt{clock\_gettime()} function, the results are returned as standard output, and the \texttt{RunAsDaemon} option needs to be disabled. The executable file has been run on all devices. The functions are used at different frequencies according to the role of the node.

\begin{figure}
    \centering
    \includegraphics[width=1\linewidth]{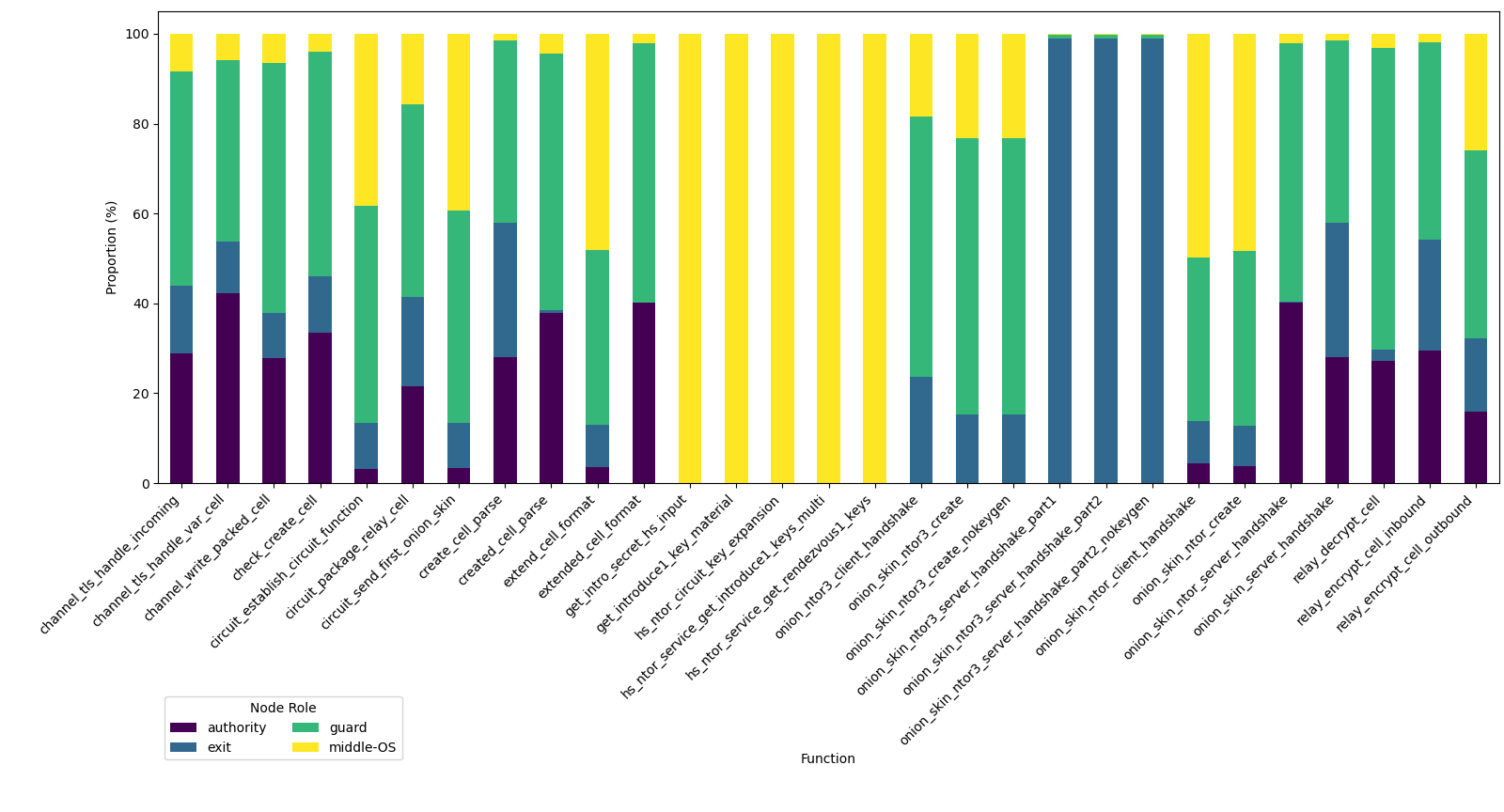}
    \caption{Proportion 
 of iterations per relay role for a given function. \textit{In this experiment, OS and middle relay are the same devices.}}
    \label{fig:funcpropo}
\end{figure}

Focusing on the traditional cryptography used, the table \ref{fig:OpenSSLbenchlocal} below presents the performance achieved by Raspberry Pi 4 and 5 (4GB) using OpenSSL.

\begin{table}
    \centering
        \caption{Benchmark of Tor's asymmetric schemes on Raspberry Pi. \textit{Average operation iteration per second.}}
    \label{fig:OpenSSLbenchlocal}
    \small
    \begin{tabularx}{\textwidth}{@{}l *{5}{>{\centering\arraybackslash}X}@{}}
        \toprule
        & \textbf{\scriptsize sign/s} & \textbf{\scriptsize verify/s} & \textbf{\scriptsize Model} \\
        \midrule
        Ed25519  & 2,939.1 & 1,327.0 & 4b \\ 
        RSA2048  & 199.2 & 7,492.9 & 4b \\
        RSA3072 & 21.3 & 1,209.6 & 4b \\ \\
        Ed25519  & 16,301.1 & 6,846.1 & 5 \\ 
        RSA2048  & 281.0 & 10,289.0 & 5 \\
        RSA3072 & 90.1 & 4,652.2 & 5 \\ 
        \midrule
        & \textbf{\scriptsize key exchange/s} &  & \textbf{\scriptsize Model} \\
        \midrule
        X25519 & 1,044.6 &  & 4b \\  
        X25519 & 5,973.5 &  & 5 \\ 
        \bottomrule
    \end{tabularx}
\end{table}

The first Tor protocol to invoke these cryptosystems is the circuit-extend. Lauer et al. \cite{Lauer2020T0RTT} measured the time of the ntor handshake execution by the user at 0.30~ms, and 0.10~ms by the OR. Yet, Schanck et al. \cite{circuitExtHandshake2015} benchmarked a 0.000527 ms total performance for ntor computation. Here, a low-level approach is taken to estimate the ntor execution time: \texttt{onion\_skin\_ntor\_create} computes the first client-side step of ntor in 0.13~ms; while the function \texttt{onion\_skin\_ntor\_client\_handshake}, which performs the ntor final client-side step was executed in 0.77~ms on average by all nodes. The server-side steps of the handshake achieved an average of 2.1~ms altogether.

In the PQC migration context, the ntor v3 handshake is expected to be the standard, as previously mentioned in section \ref{codeperfo}. It performed significantly better than the classical ntor with 0.67~ms on the client side and 0.63~ms on the server side.

\begin{figure}
    \centering
    \includegraphics[width=1\linewidth]{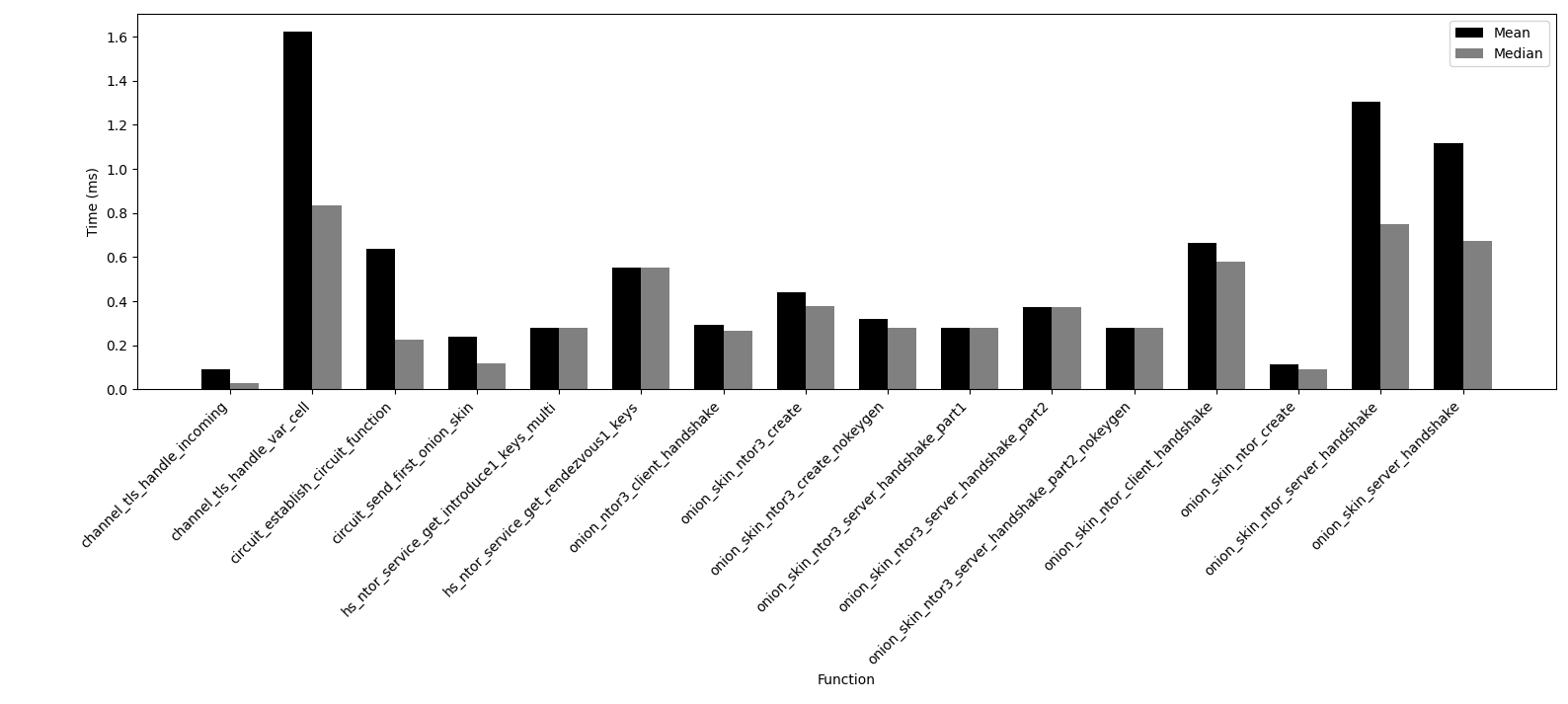}
    \caption{Execution speeds per function}
    \label{fig:functionspeed}
\end{figure}

Figure \ref{fig:functionspeed} shows the mean and median values of most measured functions; the very small metrics have been left aside for the graph's clarity.

\subsection{PQC Run Times}

The schemes of interest have been benchmarked using the OQS library to grasp the performance of the nodes against PQC schemes. Table \ref{fig:OQSbenchlocal} shows the average number of operations per second on Raspberry Pi (4GB of RAM) 4b and 5 for both KEM and signatures. Here, the Raspberry Pi 4 significantly under-performed the results published by OQS. For example, ML-KEM-512 ran 3,807 key generation operations per second, whereas Kyber-512 achieved 23,348 kengen/s \ref{fig:OQSbench}. On the other hand, Model 5 accomplished closer outputs.

\begin{table}
    \centering
        \caption{Benchmark of PQC schemes on Raspberry Pi. \textit{Average time ($\mu$s) per operation, Code retrieved from liboqs \cite{liboqs} }}
    \label{fig:OQSbenchlocal}
    \small
    \begin{tabularx}{\textwidth}{@{}l *{5}{>{\centering\arraybackslash}X}@{}}
        \toprule
        & \textbf{\scriptsize keygen/s} & \textbf{\scriptsize encaps/s} & \textbf{\scriptsize decaps/s}  & \textbf{\scriptsize Model} \\
        \midrule
        ML-KEM-512  & 3,807.3 & 3,462.0 & 2,783.0  & 4b \\ 
        ML-KEM-768 & 2,539.0 & 3,905.0 & 2,571.5  & 4b \\ 
        sntrup761  & 25.2 & 841.7 & 410.6  & 4b \\\\
        ML-KEM-512 & 23,168.3 & 19,178.0 & 15,136.3  & 5 \\ 
        ML-KEM-768 & 13,976.6 & 11,900.6 & 9,614.3  & 5 \\ 
        sntrup761  & 158.1 & 6,067.0 & 3,210.3  & 5 \\
        \midrule
        & \textbf{\scriptsize keypair/s} & \textbf{\scriptsize sign/s} & \textbf{\scriptsize verify/s}  & \textbf{\scriptsize } \\
        \midrule
        Falcon-512  & 23.6 & 615.3 & 7,866.3 & 4b \\ 
        ML-DSA-44 & 2,642.7 & 286.1 & 3,213.9  & 4b \\ \\
        Falcon-512  & 95.4 & 3,360.6 & 19,831.3  & 5 \\ 
        ML-DSA-44 & 8,986.3 & 1,885.0 & 8,139.0  & 5 \\ 
        \bottomrule
    \end{tabularx}

\end{table}

Based on the fact that a key exchange for a KEM requires key generation, encapsulation and decapsulation operations; 
\begin{center}
$T_{\text{KE}} = T_{\text{kengen}} + T_{\text{encaps}} + T_{\text{decaps}}$
\end{center}
 here, the fastest post-quantum KEM scheme performed slightly faster than X25519. The 4b model achieved 0.95~ms per classical KE exchange and 0.91~ms per ML-KEM-512 KE; nonetheless, the model 5 achieved closer results with 0.167~ms per X25519 against 0.161~ms. The operations per second are displayed in tables \ref{fig:OpenSSLbenchlocal} and \ref{fig:OQSbenchlocal}. On the client (x86\_64 architecture), X25519 was benchmarked as KEM and it resulted in 22,839.0~kegen/s, 11,950.9 encaps/s and 26,040.8~decaps/s; thus corresponding to a speed of 0.1659~ms per key exchange. Meanwhile, ML-KEM-512 produced 38,732.0 keygen/s, 33,241.3~encaps/s and 36,170~decaps/s, showing a speed difference of 66\% (between 0.1659 and 0.0835). 
From a run-time point of view, adopting ML-KEM-44 to replace Curve25519 key exchange on these devices would result in an acceleration of the overall process. Yet, this is the only scheme where X25519 is slower. On Raspberry model 4b, ML-KEM-768 showed a 14\% execution time increase while sntrup761 performed the KE operations in 43.3~ms, nearly 50 times slower than X25519. Nevertheless, on the client, the NTRU scheme was only 6 times slower than the classical key exchange with a timing of 1.005~ms/KE. Regarding DSA schemes, the Edwards-curve 25519 signing remains the fastest of the standards, followed by Falcon-512, which is still, on average, 4.8 times slower. Moreover, Ed25519 key and signature sizes (32 and 64 bytes, respectively) stay significantly smaller than both post-quantum DSA (see Tables \ref{tab:DSA} and \ref{tab:FalconDSA}), which introduces storage and transmission overhead. It should be noted that the need for flexible key management increases, as the public and private keys of Falcon and ML-DSA differ, unlike the currently used Ed25519.

In terms of verifying operation speed, Falcon slightly surpasses RSA2048 on the Raspberry Pi 4b and runs almost twice as fast in comparison to Model 5. Without taking size and key pair generation into consideration, the fastest combination of quantum-robust signature cryptosystems, as a hybrid, would be Falcon-512 with Ed25519 at security level one. RSA is not considered, as Tor is trying to move away slowly from it. However, if one considers key sizes and key pair operations per second, ML-DSA with Ed25519 is a better choice, achieving the second NIST security category.

\subsection{Reflection}

The size of the network has consequently influenced the nodes' configurations, ensuring enough guards, which has been a core challenge to achieve correct circuit building. Logs have been highly valuable alongside Nyx, a Tor command-line monitor \ref{fig:nyxoverview}.  The settings have mostly been manually deployed, and most could be easily automated with a script.

\section{Evaluation}

This section will evaluate the implementation and its outcomes in relation to the aims and objectives outlined in the initial research proposal.

\subsection{Results}
\subsubsection{Bandwidth and Circuit Performance}

\begin{figure}
    \centering
    \includegraphics[width=1\linewidth]{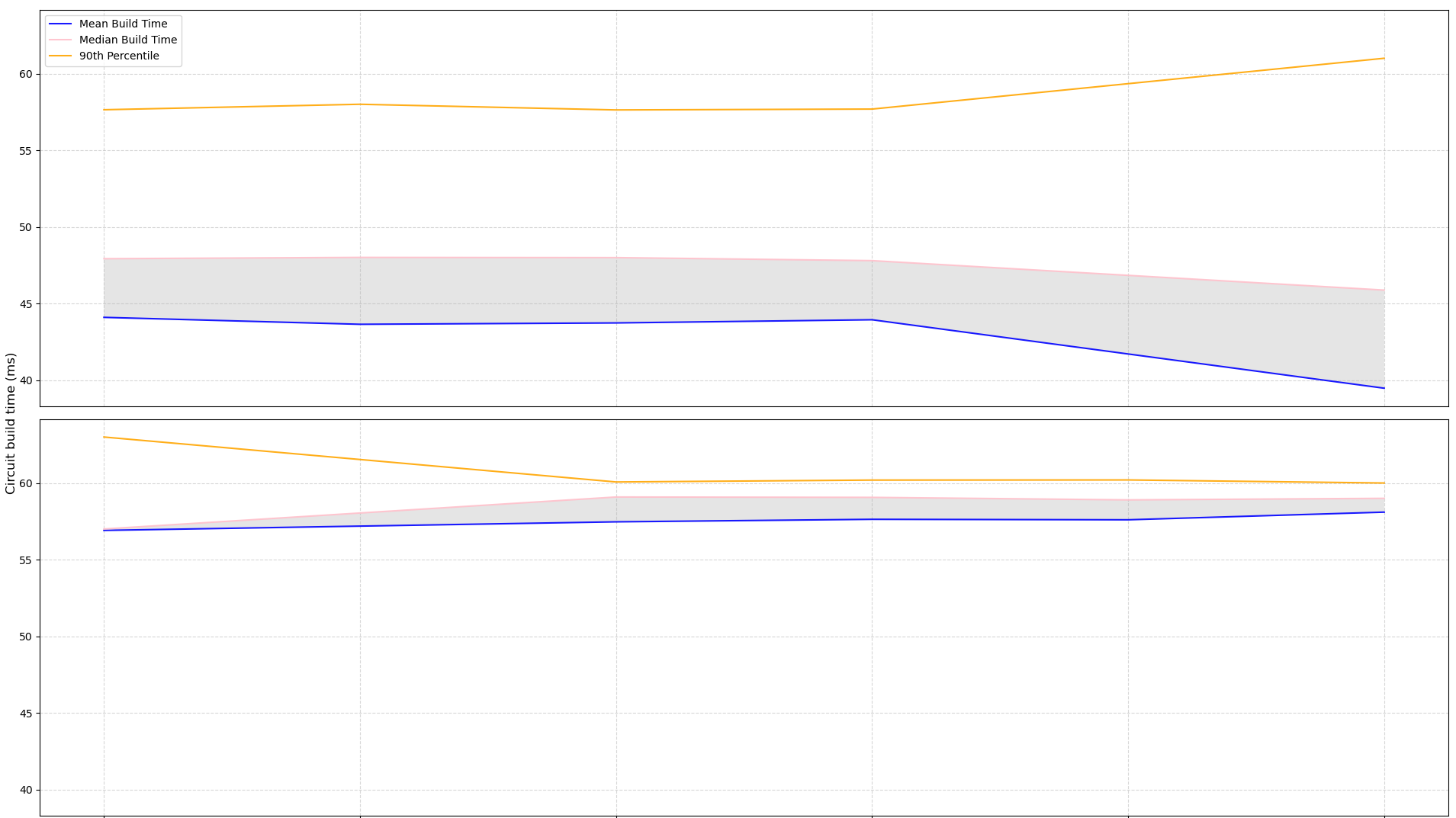}
    \caption{Circuit build time. \textit{The Q1 and Q3 measurements are omitted from the graph.}}
    \label{fig:circuitbuildtime}
\end{figure}

For an average bandwidth of 93.8 Mbits/s between nodes, without passing through the circuits in the network, the mean time taken for a circuit to be built varies between approximately 44 and 60 milliseconds. The tools used did not allow accurate measurement of the circuit build time per hop but rather the entire circuit building. The \texttt{stem} scripts, however, may be modified to measure a single-hop latency. This has not been done here, yet the overall latency gives an approximation of a single hop circuit build time. Their results are compared in Figure \ref{fig:circuitbuildtime}. The graph displays metrics from twelve different script executions over time. The upper row displays timings more spread out, yet faster than the lower one. They have been differentiated, as each script execution would either average a circuit time close to 44 ms or the upper 57~ms, due to hardware differences in the chosen nodes.

When using Tor circuits to fetch the "local.tor" HTML page (HTTPS), upload speed is 3,808.52~kB/s; download reaches 9,648.06~kB/s. Without passing through circuits, much closer values are obtained: 11,511.83~kB/s of upload speed and 11,410.24~kB/s for download speed. This implies asymmetrical bandwidth capabilities between paths; this is supported by the results obtained in the file transmission, as shown in \ref{fig:OSspeeds} with higher upload speeds on the exit path.

The Pcap analysis results have shown limitations in terms of accuracy; on the other hand, using a greater sample of curl timings data, relevant metrics are obtained.
Table \ref{fig:localTorperf} displays the median and average values of a single request according to the targeted source. The three targets are running Nginx web servers, all equipped with SSL certificates. 

\begin{table}
    \centering
     \caption{Performance of web traffic through local Tor network. \textit{Total time (ms) per curl operation.}}
    \label{fig:localTorperf}
    \small
    \begin{tabularx}{\textwidth}{@{}l *{5}{>{\centering\arraybackslash}X}@{}}
        \toprule
        & \textbf{\scriptsize Duckduckgo} & \textbf{\scriptsize Local webserver} & \textbf{\scriptsize Onion Service}  & \textbf{\scriptsize Request type} \\
        \midrule
        Mean & 262.9 & 136.7 & 174.6  & GET \\ 
        Median & 260.8 & 142.4 & 166.0  & GET \\

        Mean & \_  & 141.8 &  172.6 & POST \\ 
        Median & \_  & 148.0 & 171.5  & POST \\
        \bottomrule
        \end{tabularx}
  
\end{table}

The round-trip latency can be calculated accurately by subtracting the average latency of direct route requests from that of requests routed through Tor. Based on 1,900 requests per source, fetching "duckduckgo.com" yields an overhead of 83.1~ms, while accessing the "local.tor" Nginx webserver incurs an additional latency of 82.4 milliseconds. Rounding up these values, the estimated latency overhead is approximately 83 milliseconds.

Regarding OS performance, the data exchange to fetch the test page (as shown in Figure \ref{fig:OStest}) takes approximately 20 ms extra compared to the identically configured Nginx server. Knowing that OS requires going through three more hops than the external resource fetching it, the time per hop extension for fetching via OS is approximately 7~ms, while the circuit-building time per hop ranges from 15 to 20~ms (based on stem script results). This could be explained by the optimised ntor handshake varied initiated in \texttt{hs\_ntor.c}.

\begin{figure}
    \centering
    \includegraphics[width=1\linewidth]{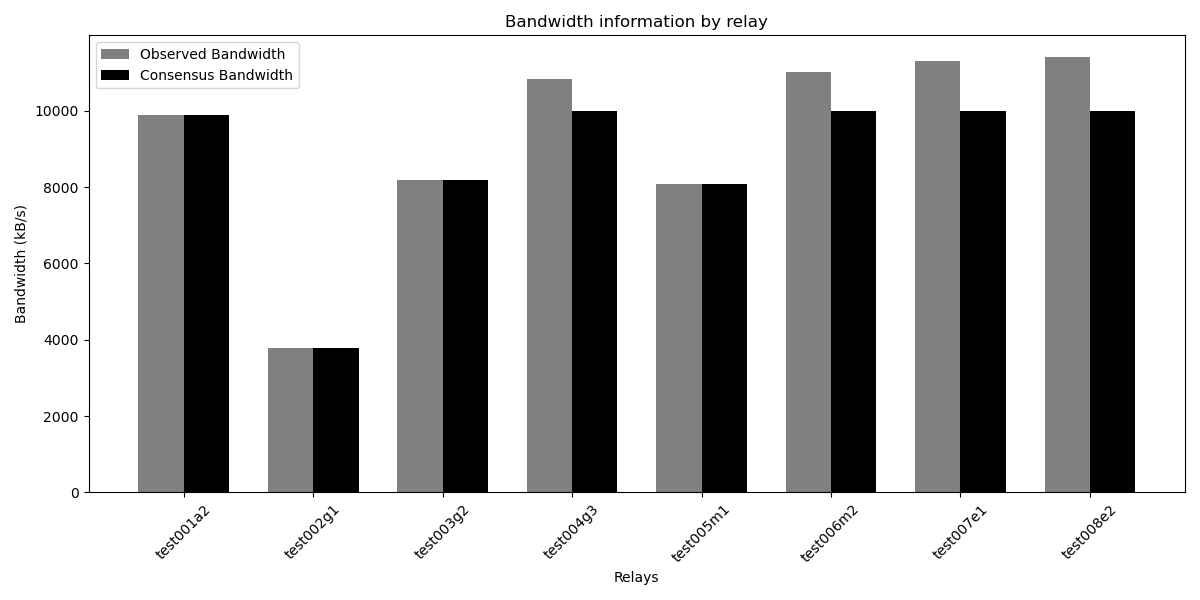}
    \caption{\texttt{sbws} upload bandwidth per node}
    \label{fig:sbwslatest}
\end{figure}

The bandwidth results obtained from \texttt{sbws} corroborate the initial bandwidth measurement of 93~Mbits/s made using iPerf. Notably, the first authority guard is excluded from this analysis as it introduces an error. In Figure \ref{fig:sbwslatest}, the first authority guard is omitted as it results in an error due to the fact that \texttt{sbws} uses it to fetch data in the network, corrupting its bandwidth assessment. 

\subsubsection{Cryptography Comparison}

The comparison potentially demonstrates that the optimisation of the X25519 implementation allows constrained devices to execute its KE almost as fast as the ML-KEM-512 standard, while on a 32GB machine, the post-quantum standard is able to perform at least 50\% faster than its classical counterpart. In a similar manner, NTRU (sntrup761) does much better on the client than it does on the Raspberry Pi. Memory usage should be compared for a deeper analysis. Falcon-512 is the fastest on the constrained devices. However, it represents a much larger latency on more powerful hardware (e.g., laptops). It should be noted that the architecture - here \texttt{aarch64} for the Raspberry Pis - should be taken into account; the OpenSSL executable for \texttt{x86\_64}, used on the laptop, provided more accurate computations of each operation per scheme.

In terms of roles, the analysis, of results displayed in Figure \ref{fig:funcpropo}, reveals distinct functional distributions among Tor nodes, highlighting the specialised responsibilities of each. Guard nodes show the largest proportion of iteration for encryption and circuit-related operations \texttt{relay\_encrypt\_cell\_inbound, relay\_decrypt\_cell, and circuit\_package\_relay\_cell}, whereas the middle node, also acting as an OS mainly displayed handshake-related function iterations. As expected, exit nodes primarily engaged in outbound traffic handling (\texttt{relay\_encrypt\_cell\_outbound} and \texttt{channel\_write\_packed\_cell}). Authority nodes have a smaller, consistent presence, reflecting their role in network management and verification rather than direct cell processing. Thus, Onion service nodes evidently exhibited the highest usage of public-key cryptography functions, followed by guard nodes, executing the most handshake-related operations.

\subsection{Effectiveness}

The network benchmarking process displays consistent results. Asymmetric latencies have been identified; however, they are not necessarily problematic. The real Tor network relies on a diversity of hardware, resulting in asymmetric bandwidth. Compared to advertised metrics (see Figure \ref{fig:qqq2}), the circuit round-trip latencies share common characteristics in their results, with a higher but faster onion latency compared to when directed to a public server. 
On the other hand, estimating a precise proportion of public key cryptography processes is difficult by solely analysing the timings of cryptographic functions.

\subsubsection{Related Work}

Zsolt et al. \cite{tujner2020qsor} highlight the impact of PQC on Tor but focus mainly on the scheme's performances and less on Tor software. Their work acknowledges the advantage of lattice-based KEM cryptosystems but presents SIKE as an optimal choice; it does not cover the efficiency of DSA schemes. In comparison, this paper benefits from a more advanced status in the NIST standardisation process. Evgnosia-Alexandra \cite{Kelesidis} gives a brief cryptographic analysis of potential candidates and emphasises hybrid Ring-LWE approaches. Yet, Evgnosia-Alexandra's paper also promotes SIKE as the appropriate choice to minimise overhead. In contrast, this work dived into Tor's software performances in a constrained environment evaluating the potential PQC load in a pragmatic manner.

\subsection{Reflection}

Based on the frequency analysis of C function execution, middle-OS and guard nodes appear to perform the most asymmetric cryptographic operations. It can be inferred that considering previous benchmarks of PQC algorithms on Raspberry Pis, their minimum hardware memory requirements may need to be increased. However, reducing the acceptable criteria could also lead to a reduction in network capacity.

\section{Conclusion}

Overall, this work has presented several methods to evaluate a local Tor network, which can be easily implemented and repeated in future testing environments. It also presented a reference point, providing results of these measurement techniques that one could consider in a similar experiment.


The key deliverables of the paper are:
\begin{itemize}
    \item Four algorithms have been demonstrated to be more appropriate than others: ML-KEM, NTRU (sntrup761) as KEMs, followed by ML-DSA and Falcon as signature schemes.
    \item A local Tor network has been successfully deployed with multiple physical devices; it reached a Tor circuit-round latency of 83~ms for a non-encapsulated bandwidth mean value of 93.8~Mbits/s.

    \item The C code related to Tor cryptographic operations has been timed.  However, the relevance of the timings should be interpreted with caution within the context of this experiment.

    \item Using the obtained results alongside PQC performances, an accurate estimate of the overhead percentage cannot be determined. Numerous unclear variables are likely to influence the final latency evolution. This work has highlighted that the hardware characteristics of Tor nodes will be one of these factors.
    
\end{itemize}

Most of the raw results have been made publicly available at \footnote{https://gitlab.torproject.org/diunisu/testing-network-results}. 

\subsection{Limitations}

The drawbacks of this work can be represented by three main points below. 

\begin{itemize}
    \item While trying to satisfy the low-bandwidth requirement, the selected schemes only cover the two first NIST security categories. Furthermore, they lack diversity in their mathematical assumptions, relying solely on lattice problems.
    
    \item The benchmarking is heavily subjective. \\ Firstly, it is influenced by the technology used, such as the optimisation of cryptosystems based on a CPU architecture. In this context, this theoretical bare-metal Tor network simulation could be improved by using hardware generally used in the real Tor network. This would require more intrusive relay metrics and may go against security and privacy principles. 
    Secondly, the timings are greatly influenced by the configuration. The torrc files have been adapted to obtain a functional local network with the given number of nodes. Yet, another blend of settings could result in a more realistic implementation. Adding more nodes should improve the stability of the setup.
    Going further, the nodes could be distanced geographically. Planet Lab \cite{PlanetLabEurope} may be an appropriate test bed.
    Additionally, the methods and tools employed in the measurement process can significantly alter the results, for instance time is consistently taken as reference points and other measuring units such as CPU cycles are neglected. 

    \item The estimation of the PQC overhead is only theoretical and misses a multitude of parameters, such as different traffic types or volumes. In a virtual environment, further testing with Shadow could provide valuable insights based on the real Tor traffic, while on physical machines, adapting or creating tools like OnionPerf to target local instantiations could also produce more relevant measures. In the future, Tor's software optimisations and restructuring of the underlying protocols will also impact network performance.
\end{itemize}

\subsection{Future Work}

Future work includes advancing the post-quantum circuit-extend handshake design, inter alia, by comparing a DSA-free handshake to a KEM authentication (still requiring DSA for static leaf signature). This should be followed by a software implementation draft to enable practical evaluation of the integration of these PQC schemes. It would be particularly useful to assess its performance in a realistic scenario. Looking ahead, when the Arti code will allow the running of relays, it will be helpful to note the difference in cryptography performance between C Tor and Rust in this context.


\newpage

\section*{Glossary}

\addcontentsline{toc}{section}{Glossary} 

\begin{center}
\begin{longtable}{m{3cm}l}
\textbf{Acronym} & \textbf{Meaning} \\ 
\hline
\\
\hypertarget{ACCE}{ACCE} & Authenticated and Confidential Channel Establishment\\
\hypertarget{ACME}{ACME} & Automated Certificate Management Environment\\
\hypertarget{AEAD}{AEAD} & Authenticated Encryption with Associated Data\\
\hypertarget{AES}{AES} & Advanced Encryption Standard\\
\hypertarget{AKE}{AKE} & Authenticated Key Exchange\\
\hypertarget{ARTI}{ARTI} & A pure-Rust Tor Implementation\\
\hypertarget{CA}{CA} & Certificate Authorities\\
\hypertarget{CCA}{CCA} & Chosen-Ciphertext Attacks\\
\hypertarget{SVP}{SVP} & Shortest Vector Problem\\
\hypertarget{CRYSTALS}{CRYSTALS} & Cryptographic Suite for Algebraic Lattices\\
\hypertarget{DH}{DH} & Diffie-Hellman \\
\hypertarget{DHCP}{DHCP} & Dynamic Host Configuration Protocol \\
\hypertarget{DNS}{DNS} & Domain Name System\\
\hypertarget{DSA}{DSA} & Digital Signature Algorithm \\
\hypertarget{ECC}{ECC} & Elliptic Curve Cryptography\\
\hypertarget{ECDH}{ECDH} & Elliptic-Curve Diffie-Hellman\\
\hypertarget{EES}{EES} & Efficient Embedded Security\\
\hypertarget{FALCON}{FALCON} & Fast Fourier Lattice-based Compact signatures over NTRU\\
\hypertarget{FO}{FO} & Fujisaki-Okamoto transform\\
\hypertarget{FS}{FS} & Forward Secrecy\\
\hypertarget{HHK}{HHK} & Hofheinz-Hövelmanns-Kiltz transform\\
\hypertarget{HQC}{HQC} & Hamming Quasi-Cyclic\\
\hypertarget{IETF}{IETF} & Internet Engineering Task Force\\
\hypertarget{IND-CCA}{IND-CCA} & INDistinguishability under Chosen-Ciphertext Attack\\
\hypertarget{IND-CCA2}{IND-CCA2} & INDistinguishability under Adaptive-Ciphertext Attack\\
\hypertarget{IND-CPA}{IND-CPA} & INDistinguishability under Chosen-Plaintext Attack\\
\hypertarget{IP}{IP} & Internet Protocol\\
\hypertarget{KEM}{KEM} & Key Encapsulation Mechanism\\
\hypertarget{LMS}{LMS} & Leighton–Micali Signature\\
\hypertarget{MLWE}{MLWE} & Module Learning With Error\\
\hypertarget{ML-KEM}{ML-KEM} & Module-Lattice-based Key-Encapsulation Mechanism\\
\hypertarget{ML-DSA}{ML-DSA} & Module-Lattice-based Digital Signature Algorithm\\
\hypertarget{MTU}{MTU} & Maximum Transmission Unit\\
\hypertarget{NAT}{NAT} & Network Address Translation\\
\hypertarget{NIST}{NIST} & National Institute of Standards and Technology\\
\hypertarget{NSA}{NSA} & National Security Agency\\
\hypertarget{NTRU}{NTRU} & Nth degree TRUncated polynomial ring\\
\hypertarget{OS}{OS} & Onion Services\\
\hypertarget{PDK}{PDK} & Pre-Distributed (public) Keys\\
\hypertarget{PFS}{PFS} & Perfect Forward Secrecy\\
\hypertarget{PKE}{PKE} & Public Key Encryption\\
\hypertarget{PQC}{PQC} & Post-Quantum Cryptography\\
\hypertarget{PRNG}{PRNG} & Pseudo-Random Number Generator\\
\hypertarget{QC}{QC} & Quantum Computer\\
\hypertarget{QKD}{QKD} & Quantum Key Distribution\\
\hypertarget{RAM}{RAM} & Random-Access Memory\\

\textbf{Acronym} & \textbf{Meaning} \\
\\
\hypertarget{RSA}{RSA} & Rivest-Shamir-Adleman\\
\hypertarget{RTT}{RTT} & Round-Trip Time\\
\hypertarget{SHA}{SHA} & Secure Hash Algorithms\\
\hypertarget{SSL}{SSL} & Secure Sockets Layer\\
\hypertarget{TAP}{TAP} & Tor Authentication Protocol\\
\hypertarget{TCP}{TCP} & Transmission Control Protocol\\
\hypertarget{TLS}{TLS} & Transport Layer Security\\
\hypertarget{TOR}{TOR} & The Onion Routing\\
\hypertarget{XMSS}{XMSS} & eXtended Merkle Signature Scheme\\
\hypertarget{1W-AKE}{1W-AKE} & One-Way Authenticated Key Exchange\\
\\
\hline
\end{longtable}
\end{center}



\appendix
\section[\appendixname~\thesection]{}\label{tormetrics}


\begin{figure}[H]
    \centering
    \includegraphics[width=0.8\linewidth]{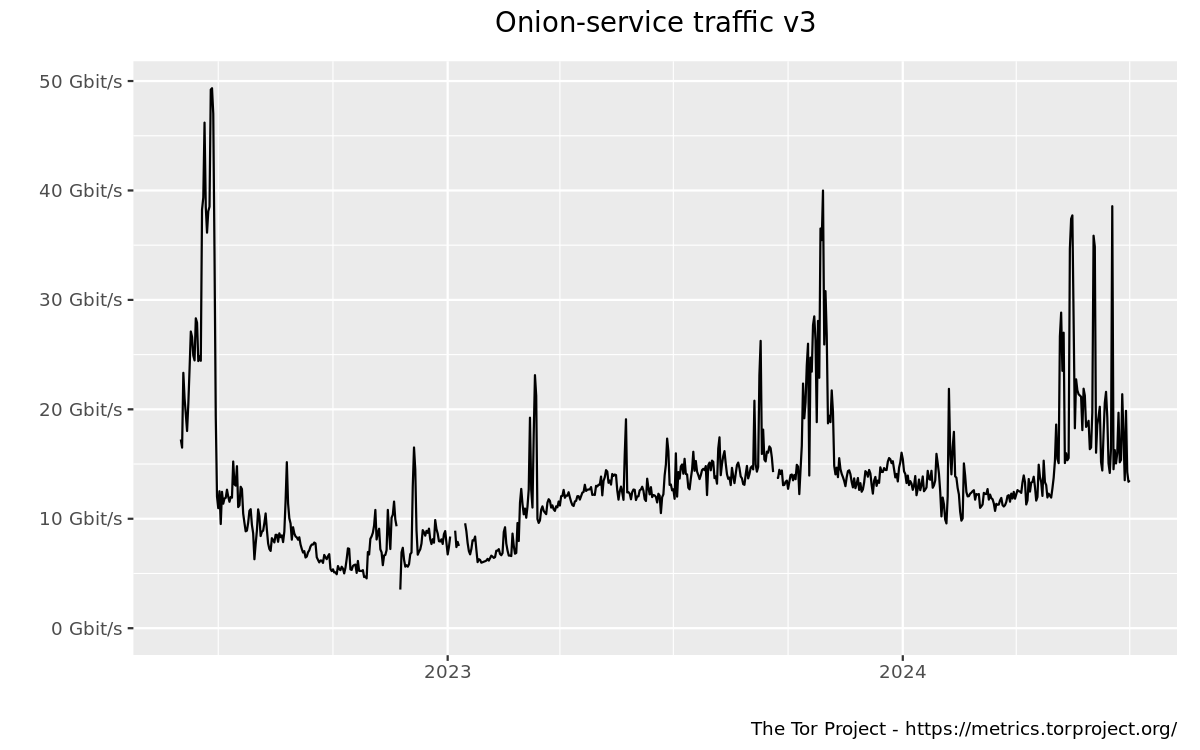}
    \caption{V3 onion service traffic over 2023}
    \label{fig:v3Onion}
\end{figure}

\begin{figure}[H]
    \centering
    \includegraphics[width=0.8\linewidth]{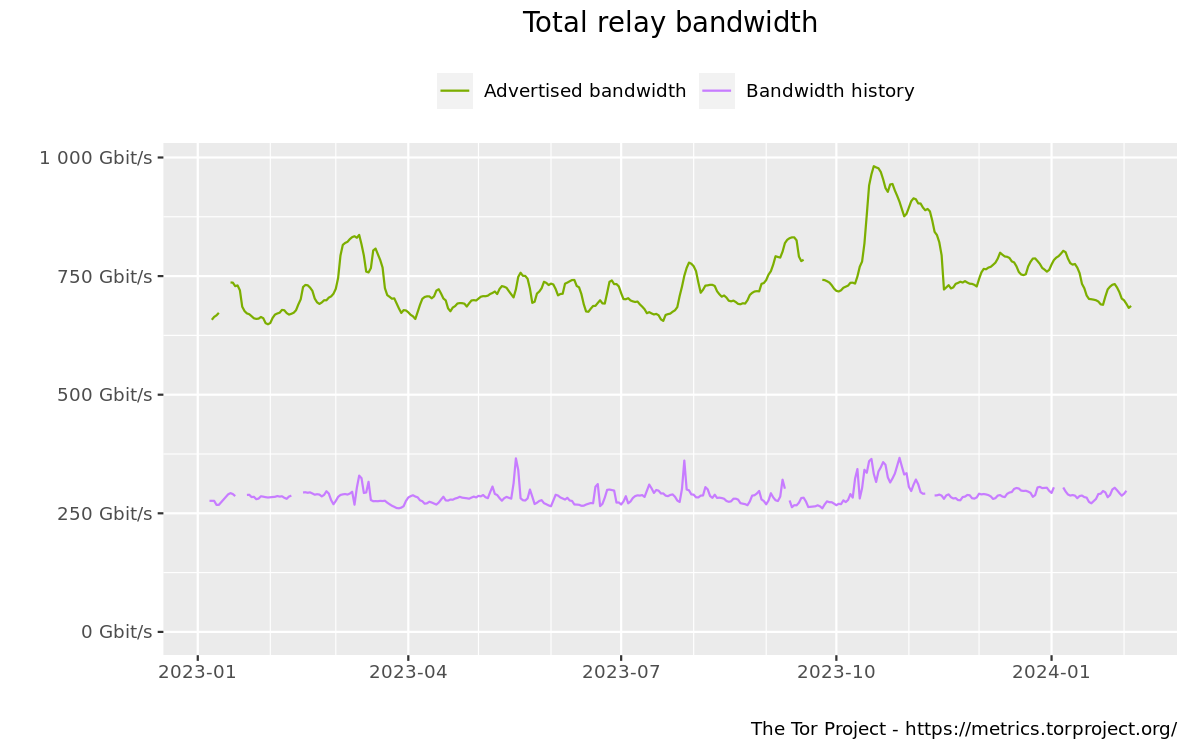}
    \caption{Total bandwidth traffic between over 2023}
    \label{fig:v3Onion2} 
\end{figure}

\begin{figure}
    \centering
    \includegraphics[width=1\linewidth]{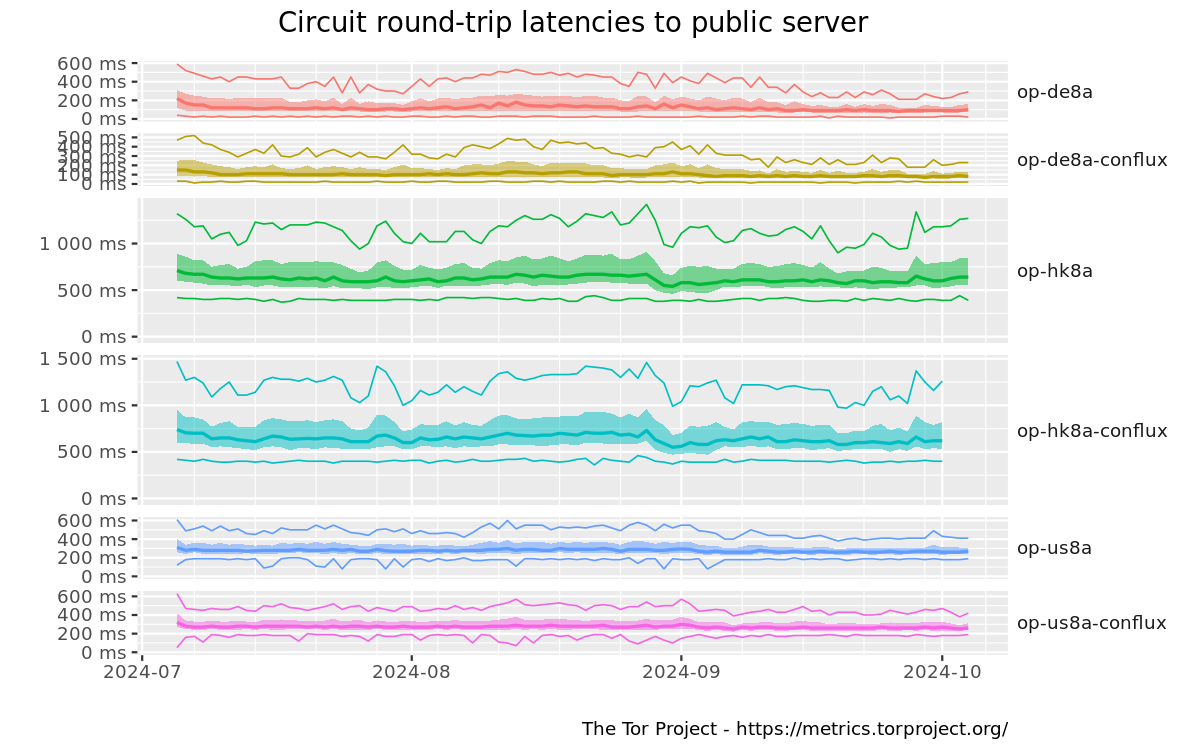}
    \vspace{0.5cm} 
    \includegraphics[width=1\linewidth]{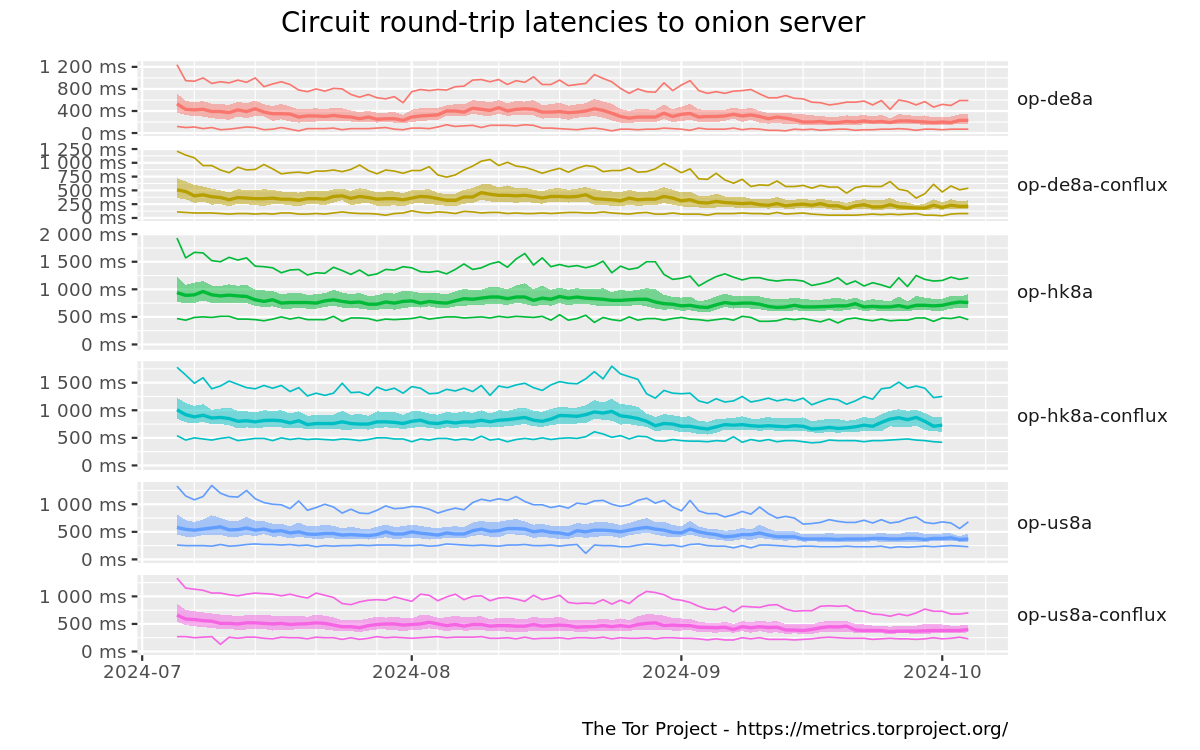}
    \caption{OnionPerf latencies between July 2024 and September 2024}
    \label{fig:qqq2} 
\end{figure}

\begin{table}
    \centering
        \caption{Relay Keys. \textit{Source: Tor gitlab \cite{torspec}}} \label{tab:relayKeys}
    \includegraphics[width=1\linewidth]{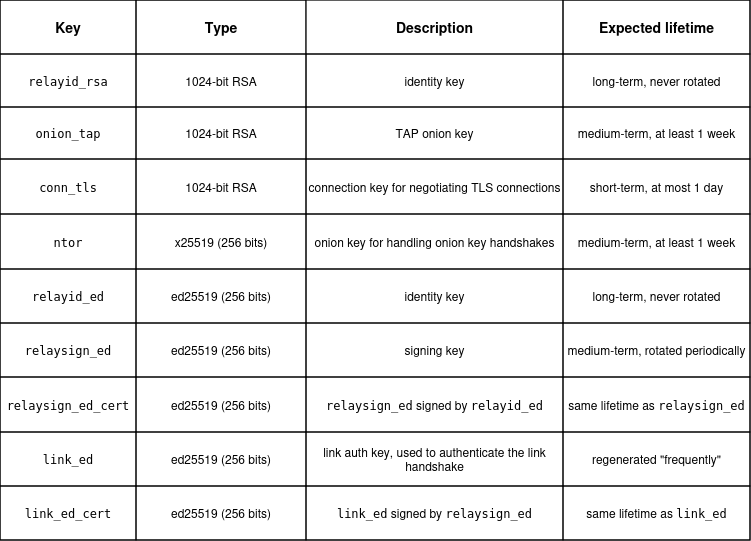}
\end{table}

\begin{table}
    \centering
       \caption{Onion Service Keys. \textit{Source: Tor gitlab \cite{torspec}}} \label{tab:OnionServiceKeys}
    \includegraphics[width=1\linewidth]{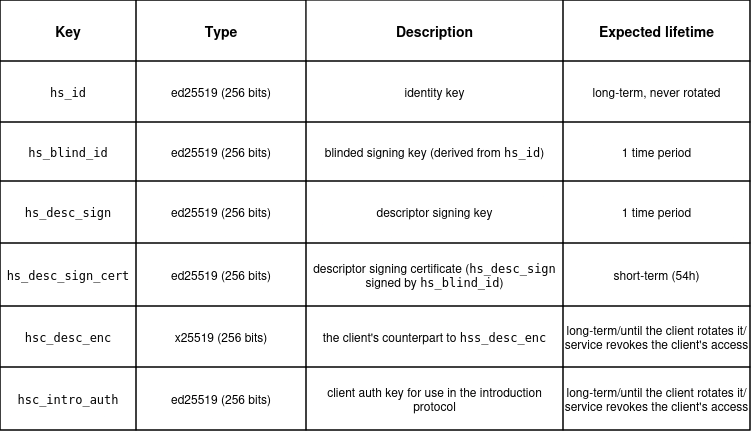}
\end{table}

\begin{table}
    \centering
       \caption{Directory authorities Keys. \textit{Source: Tor Gitlab \cite{torspec} and /var/lib/tor/cached-certs}}  \label{tab:DSA}
    \includegraphics[width=1\linewidth]{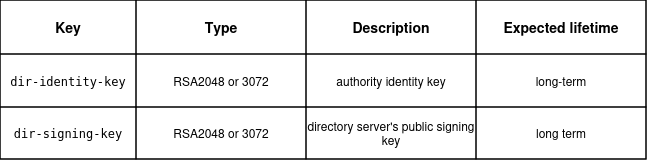}
\end{table}

\FloatBarrier

\begin{mdframed}\label{Scheme specification} 
\begin{itemize}
    \item \textbf{AES-128-CTR}: in counter mode as stream cipher; IV of all 0 bytes.
    \item \textbf{RSA}: exponent of 65,537. OAEP-MGF1 padding, with SHA-1 as its digest function (label left unset). Tor paper plans to move away from SHA-1 to align with evolving cryptographic standards and best practices.

    \item \textbf{SHA-1} as "hash of a public key", DER encoding of an ASN.1 RSA public key (as specified in PKCS.1).    
    \item  \textbf{Relay identity keys}: currently two types are still deployed: RSA1024 is legacy and being replaced by Ed25519 which is the one advertised in the table.
\item \textbf{Tor's TLS 1.2 Ciphers:}
    \begin{itemize}
        \item TLS1\_TXT\_DHE\_RSA\_WITH\_AES \_256\_SHA
        \item TLS1\_TXT\_DHE\_RSA\_WITH\_AES \_128\_SHA
        \item TLS1\_TXT\_ECDHE\_RSA\_WITH\_ AES\_256\_GCM\_SHA384
        \item TLS1\_TXT\_ECDHE\_RSA\_WITH\_ AES\_128\_GCM\_SHA256
        \item TLS1\_TXT\_ECDHE\_RSA\_WITH\_ AES\_256\_CBC\_SHA
        \item TLS1\_TXT\_ECDHE\_RSA\_WITH\_ AES\_128\_CBC\_SHA
    \end{itemize}

\item \textbf{Tor's TLS 1.3 AEAD Ciphers:}
    \begin{itemize}
        \item TLS1\_3\_TXT\_AES\_128\_GCM \_SHA256
        \item TLS1\_3\_TXT\_AES\_256\_GCM \_SHA384
        \item TLS1\_3\_TXT\_CHACHA20\_POLY1305 \_SHA256
        \item TLS1\_3\_TXT\_AES\_128\_ CCM\_SHA256
    \end{itemize}

\end{itemize}
\end{mdframed}

\captionsetup{justification=centering} 
\captionof{figure}{Schemes specification for Table \ref{tab:tableschemes}} \caption*{\small \textit{Source: Tor's gitlab (\texttt{tortls\_openssl.c}, \texttt{tortls\_nss.c}, \texttt{tortls.c}), Specifications \cite{torspec}}}

\begin{table}
    \centering
       \caption{DSA standards released by NIST in 2024. \textit{Source: FIPS 204, 205\cite{FIPS204}; Size displayed in bytes}} \label{tab:DSA}
    \noindent\makebox[\textwidth]{%
    \includegraphics[width=1\linewidth]{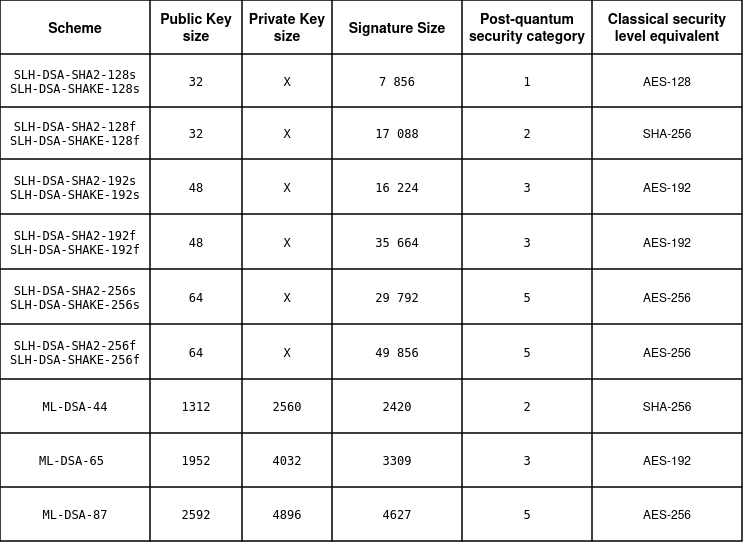}}
\end{table}

\begin{table}
    \centering
      \caption{Hashed-based DSA standardised by NIST in 2020.  \textit{Source: Open Quantum Safe \cite{liboqs}; Size displayed in bytes, four example settings selected among others}}\label{tab:DSA2020}
    \includegraphics[width=1\linewidth]{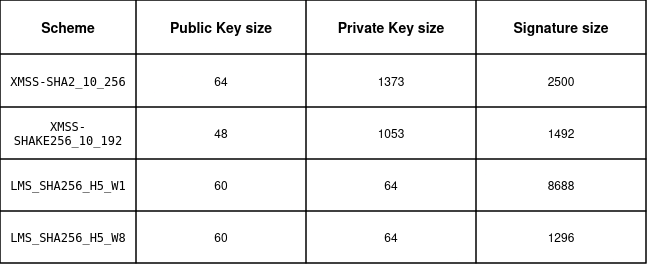}
\end{table}

\begin{table}
    \centering
       \caption{Falcon DSA key sizes.  \textit{Source: Open Quantum Safe \cite{liboqs};  Size displayed in bytes}}   \label{tab:FalconDSA}
    \includegraphics[width=1\linewidth]{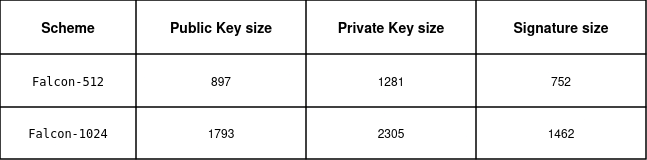}
\end{table}

\label{networkarchitecture}
\begin{figure}
    \includegraphics[width=0.8\linewidth]{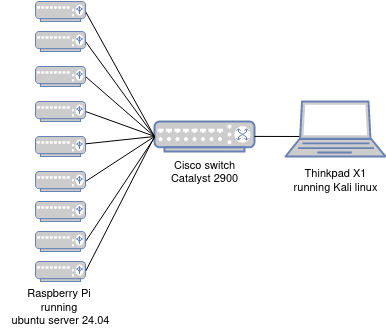}
          \caption{Hardware used for the implementation}
    \label{fig:hardware}
\end{figure}

\label{configuration} 
\lstset{
  basicstyle=\scriptsize\ttfamily, 
  breaklines=true,                 
  frame=single,                    
  language=bash,                   
  showstringspaces=false
}
\begin{lstlisting}[caption=Cisco switch configuration, label=lst:ciscoconf, float=hbt!]
> enable
configure terminal
ip dhcp pool LAN_POOL
network 192.168.1.0 255.255.255.0
interface vlan1
ip address 192.168.1.1 255.255.255.0
no shutdown
service dhcp
exit
write memory
\end{lstlisting}

\begin{lstlisting}[caption=Chutney configuration file, label=lst:chutconf, float=hbt!]
Authority1 = Node(tag="a1", authority=1, relay=1, torrc="authority.tmpl")
Authority2 = Node(tag="a2", authority=1, relay=1, torrc="authority.tmpl")

GuardRelay1 = Node(tag="g1", relay=1, guard=1, torrc="relay.tmpl")
GuardRelay2 = Node(tag="g2", relay=1, guard=1, torrc="relay.tmpl")
GuardRelay3 = Node(tag="g3", relay=1, guard=1, torrc="relay.tmpl")

MiddleRelay1 = Node(tag="m1", relay=1, middle=1, onion_service=1, torrc="relay.tmpl")  
MiddleRelay2 = Node(tag="m2", relay=1, middle=1, rendezvous=1, torrc="relay.tmpl")  

ExitRelay1 = Node(tag="e1", relay=1, exit=1, torrc="relay.tmpl")
ExitRelay2 = Node(tag="e2", relay=1, exit=1, torrc="relay.tmpl")

NODES = [Authority1, Authority2, GuardRelay1, GuardRelay2, GuardRelay3, MiddleRelay1, MiddleRelay2, ExitRelay1, ExitRelay2]

ConfigureNodes(NODES)
\end{lstlisting}

\begin{lstlisting}[caption=Listing of files generated by node type (chutney), label=lst:chutfiles, float=hbt!]
000a1 (Directory authority)
  fingerprint
  fingerprint-Ed25519
  keys
  authority_certificate
  authority_identity_key
  authority_signing_key
  Ed25519_master_id_public_key
  Ed25519_master_id_secret_key
  Ed25519_signing_cert
  Ed25519_signing_secret_key
  secret_id_key
  secret_onion_key
  secret_onion_key_ntor
  lock
  torrc
  
004g3 (Guard relay)
  fingerprint
  fingerprint-Ed25519
  keys
  Ed25519_master_id_public_key
  Ed25519_master_id_secret_key
  Ed25519_signing_cert
  Ed25519_signing_secret_key
  secret_id_key
  secret_onion_key
  secret_onion_key_ntor
  lock
  torrc
 
005m1 (Middle relay)
  fingerprint
  fingerprint-Ed25519
  keys
  Ed25519_master_id_public_key
  Ed25519_master_id_secret_key
  Ed25519_signing_cert
  Ed25519_signing_secret_key
  secret_id_key
  secret_onion_key
  secret_onion_key_ntor
  lock
  torrc

007e1 (Exit relay)
  fingerprint
  fingerprint-Ed25519
  keys
  Ed25519_master_id_public_key
  Ed25519_master_id_secret_key
  Ed25519_signing_cert
  Ed25519_signing_secret_key
  secret_id_key
  secret_onion_key
  secret_onion_key_ntor
  lock
  torrc
\end{lstlisting}

\begin{lstlisting}[caption=Directory authority torrc file, label=lst:authtorrc, float=hbt!]
TestingTorNetwork 1

PathsNeededToBuildCircuits 0.35
TestingDirAuthVoteExit $D36D51189EC2112507DDB293FC166C5E5ADA9B91,$3E5729E0BE77AAD422BE2D95D72200EF01D1F29F
TestingDirAuthVoteExitIsStrict 1
TestingDirAuthVoteHSDir $BD452B626D1A453BB3E557B363CEC39F1E63D089,$CCE340CBBFFA07C8E434261FFCA5AB457DD06217,$D36D51189EC2112507DDB293FC166C5E5ADA9B91
#TestingDirAuthVoteHSDirIsStrict 1
TestingDirAuthVoteGuard $84C8BDC0ECB2A6F7B536A2C6696DE0ADE8AC09A9,$139726CAC0132D62C1D6E5E2F306315CCFE857A9,$A0595788A046B7B2DF48B04A02C262DC525A82E2,$691B61F16F6D48F65183AF7A39BC789645C14C49,$2EEB5B771A2A177EC0D184D410DDB4F153FF354F
TestingDirAuthVoteGuardIsStrict 1
V3AuthNIntervalsValid 2
V3BandwidthsFile /home/ubuntu/000a1/bwfile.v3bw

MiddleNodes $691B61F16F6D48F65183AF7A39BC789645C14C49,$CCE340CBBFFA07C8E434261FFCA5AB457DD06217
DataDirectory /home/ubuntu/000a1
RunAsDaemon 1
ConnLimit 60
Nickname test000a1
ShutdownWaitLength 2
DisableDebuggerAttachment 0

AddressDisableIPv6 1
ControlPort 8000
ControlSocket /home/ubuntu/000a1/control
CookieAuthentication 1
PidFile /home/ubuntu/000a1/pid

Log notice file /home/ubuntu/000a1/notice.log
Log info file /home/ubuntu/000a1/info.log
Log debug file debug.log
ProtocolWarnings 1
SafeLogging 0
LogTimeGranularity 1

AuthoritativeDirectory 1
V3AuthoritativeDirectory 1
ContactInfo auth0@test.test
DirAuthority test000a1 orport=5100 no-v2 v3ident=4832D1130A4F32E4410E42D33C36B6B87EC37C27 192.168.1.15:7100 EA30E583EF155930ACE4144D0CFA800725C65D46
DirAuthority test001a2 orport=5101 no-v2 v3ident=43DE83E17AB616F71A36394D519732767B2CEF50 192.168.1.16:7101 BD452B626D1A453BB3E557B363CEC39F1E63D089
SocksPort 9050
OrPort 5100
Address 192.168.1.15
DirPort 7100
ExitRelay 0
DisableNetwork 0
ConfluxEnabled 0
TestingAuthKeyLifetime 3 months
ServerDNSAllowNonRFC953Hostnames 1
ServerDNSDetectHijacking 0
\end{lstlisting}

\begin{lstlisting}[caption=Guard relay torrc file, label=lst:authtorrc2, float=hbt!]
TestingTorNetwork 1
PathsNeededToBuildCircuits 0.45
TestingDirAuthVoteExit *
TestingDirAuthVoteHSDir *
V3AuthNIntervalsValid 2

TestingDirAuthVoteGuard *
TestingMinExitFlagThreshold 0

DataDirectory /home/ubuntu/002g1
RunAsDaemon 1
ConnLimit 60
Nickname test002g1
ShutdownWaitLength 2
DisableDebuggerAttachment 0
AddressDisableIPv6 1
ControlPort 8002
ControlSocket /home/ubuntu/002g1/control
CookieAuthentication 1
PidFile /home/ubuntu/002g1/pid
Log notice file /home/ubuntu/002g1/notice.log
Log info file /home/ubuntu/002g1/info.log
ProtocolWarnings 1
SafeLogging 0
LogTimeGranularity 1

DirAuthority test000a1 orport=5100 no-v2 v3ident=4832D1130A4F32E4410E42D33C36B6B87EC37C27 192.168.1.15:7100 EA30E583EF155930ACE4144D0CFA800725C65D46
DirAuthority test001a2 orport=5101 no-v2 v3ident=43DE83E17AB616F71A36394D519732767B2CEF50 192.168.1.16:7101 BD452B626D1A453BB3E557B363CEC39F1E63D089

OrPort 5102
Address 192.168.1.17
ExitRelay 0
DisableNetwork 0
ConfluxEnabled 0
\end{lstlisting}

\begin{lstlisting}[caption=Client torrc file, label=lst:clienttorrc, float=hbt!]
TestingTorNetwork 1

DirAuthority test000a1 orport=5100 no-v2 v3ident=4832D1130A4F32E4410E42D33C36B6B87EC37C27 192.168.1.15:7100 EA30E583EF155930ACE4144D0CFA800725C65D46
DirAuthority test001a2 orport=5101 no-v2 v3ident=43DE83E17AB616F71A36394D519732767B2CEF50 192.168.1.16:7101 BD452B626D1A453BB3E557B363CEC39F1E63D089
SocksPort 9050
DisableNetwork 0
CircuitBuildTimeout 60
LearnCircuitBuildTimeout 0
DataDirectory /var/lib/tor
LogTimeGranularity 1

NumEntryGuards 1
UseEntryGuards 0
ClientOnly 1

SocksPolicy accept 192.168.0.0/16
RunAsDaemon 1
ConfluxEnabled 0
\end{lstlisting}

\begin{lstlisting}[caption=Tor Browser command, label=lst:TorBrowsercommand, float=hbt!]
$ TOR_SKIP_LAUNCH=1 TOR_SOCKS_PORT=9050 TOR_SKIP_CONTROLPORTTEST=1 /home/$USERNAME/.local/share/torbrowser/tbb/x86_64/tor-browser/Browser/start-tor-browser
\end{lstlisting}

\begin{figure}
    \centering
    \includegraphics[width=1\linewidth]{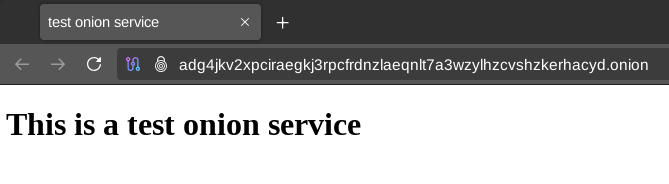}
    \caption{Onion Service test}
    \label{fig:OStest}
\end{figure}

\lstset{
  basicstyle=\scriptsize\ttfamily, 
  breaklines=true,                 
  frame=single,                    
  language=bash,                   
  showstringspaces=false
}

\begin{lstlisting}[caption=Tor Browser command, label=lst:curl3, float=hbt!]
for i in {1..500}; do                             
    curl --socks5-hostname 127.0.0.1:9050 -w "%{time_total}, " -o /dev/null -X POST \
-H "Content-Type: application/json" \
-H "User-Agent: BenchmarkingAgent/1.0" \
-H "Accept: application/json" \
-d '{
  "cookie": "value111111111111111111111111",
  "username": "testuser",
  "password": "76478736456728374637823764637382",
  "session_id": "abc123xyz4567890",
  "data": {
    "field1": "Some random text to increase payload size 1",
    "field2": "Some random text to increase payload size 2",
    "field3": "Some random text to increase payload size 3"
  }
}' \
-s https://local.tor/postpath
done
\end{lstlisting}

\begin{figure}
    \includegraphics[width=0.6\linewidth]{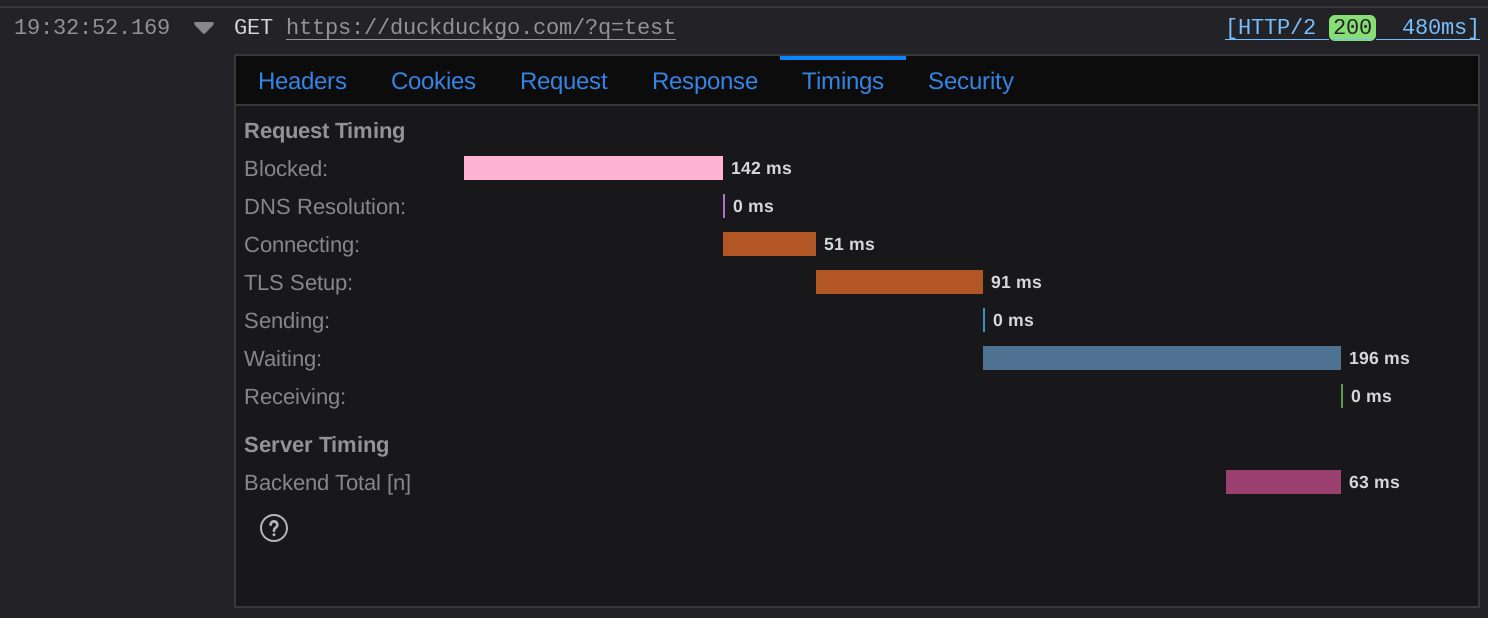}\\
    \vspace{0.5cm} 
    \includegraphics[width=0.6\linewidth]{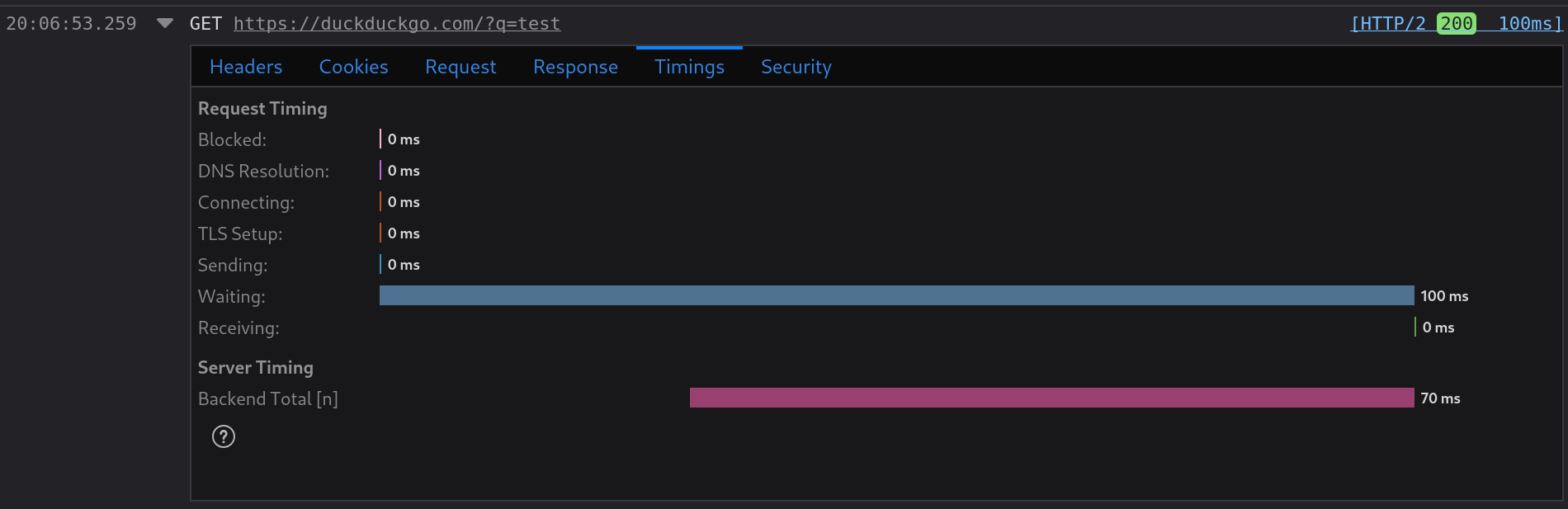}
    \caption{Comparison of request timing between the local Tor network (upper image) and direct internet access (lower image). \textit{The upper screenshot is from the Tor browser, while the lower one is from the Firefox browser.} }
    \label{fig:timingcomparison}
\end{figure}

\vspace{15pt}
\begin{figure}
    \centering
    \includegraphics[width=0.4\linewidth]{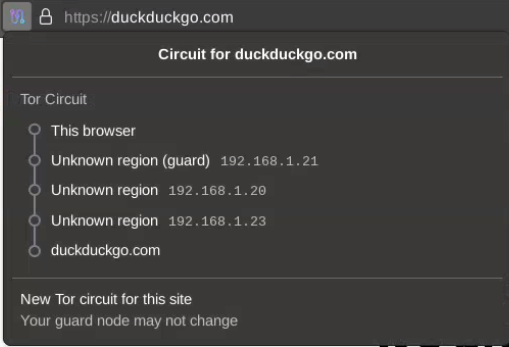}
    \caption{Tor browser displaying circuit}
    \label{fig:circuitOnBrowser}
\end{figure}

\begin{figure}
    \centering
    \includegraphics[width=1\linewidth]{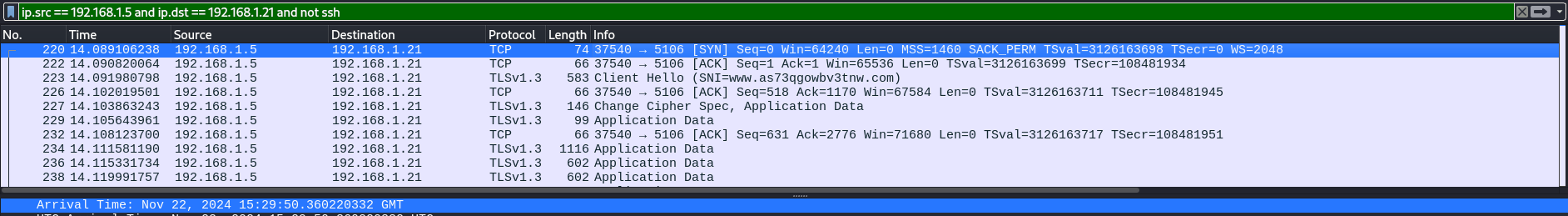}
    \vspace{0.5cm} 
    \includegraphics[width=1\linewidth]{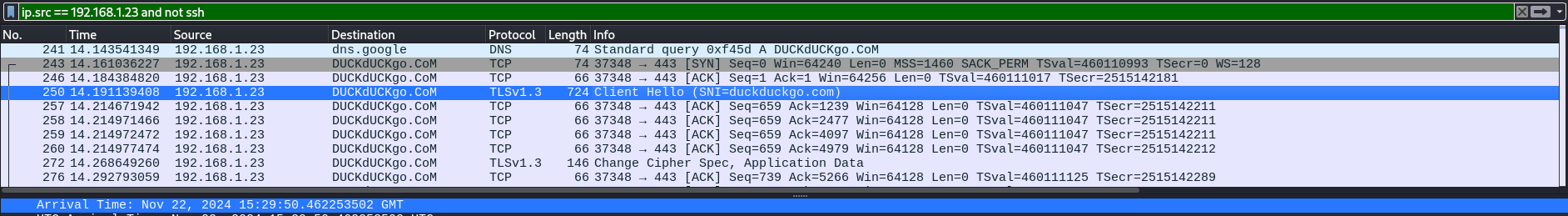}
    \caption{Wireshark sniffing eth0 interface}
    \label{fig:sniffing}
\end{figure}

\begin{figure}
    \centering
    \includegraphics[width=1\linewidth]{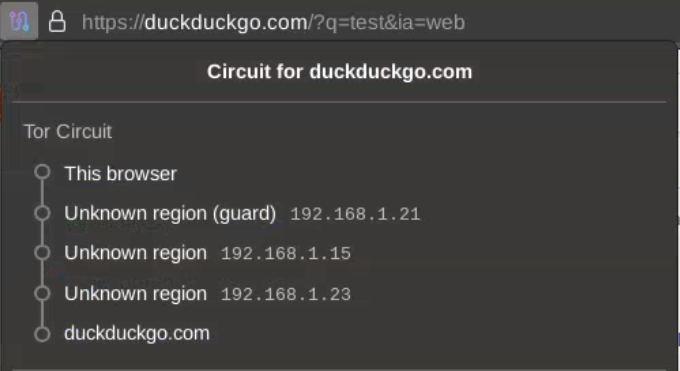}
    \caption{Tor browser displaying circuit}
    \label{fig:circuitOnBrowser2}
\end{figure}

\begin{figure}
    \centering
    \includegraphics[width=1\linewidth]{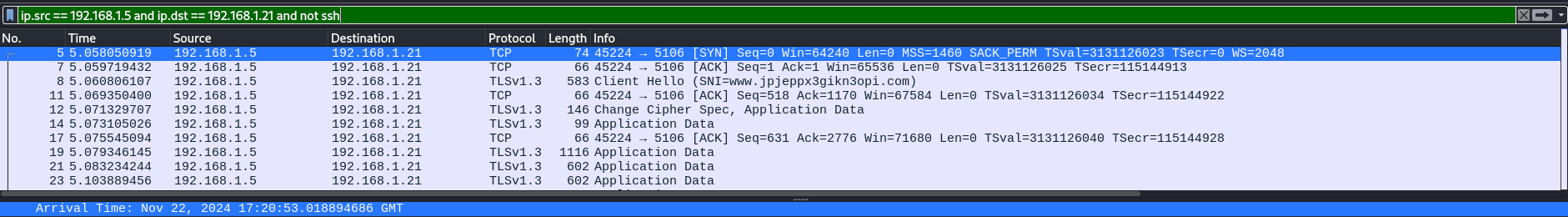}
    \vspace{0.5cm} 
    \includegraphics[width=1\linewidth]{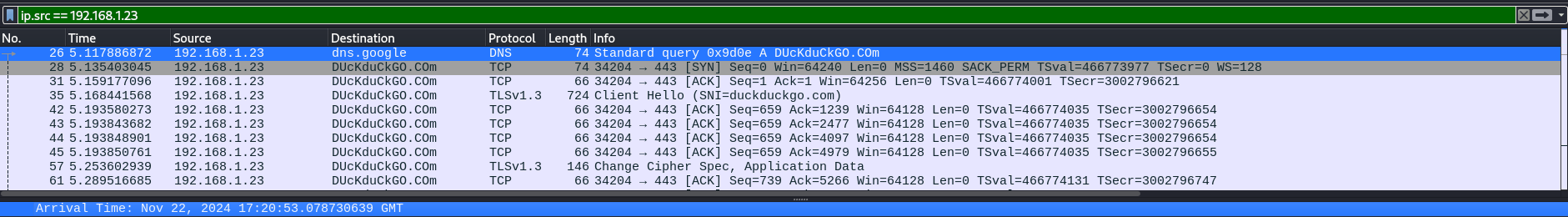}
    \caption{Wireshark sniffing eth0 interface}
    \label{fig:sniffing2}
\end{figure}

\begin{figure}
    \centering
    \includegraphics[width=1\linewidth]{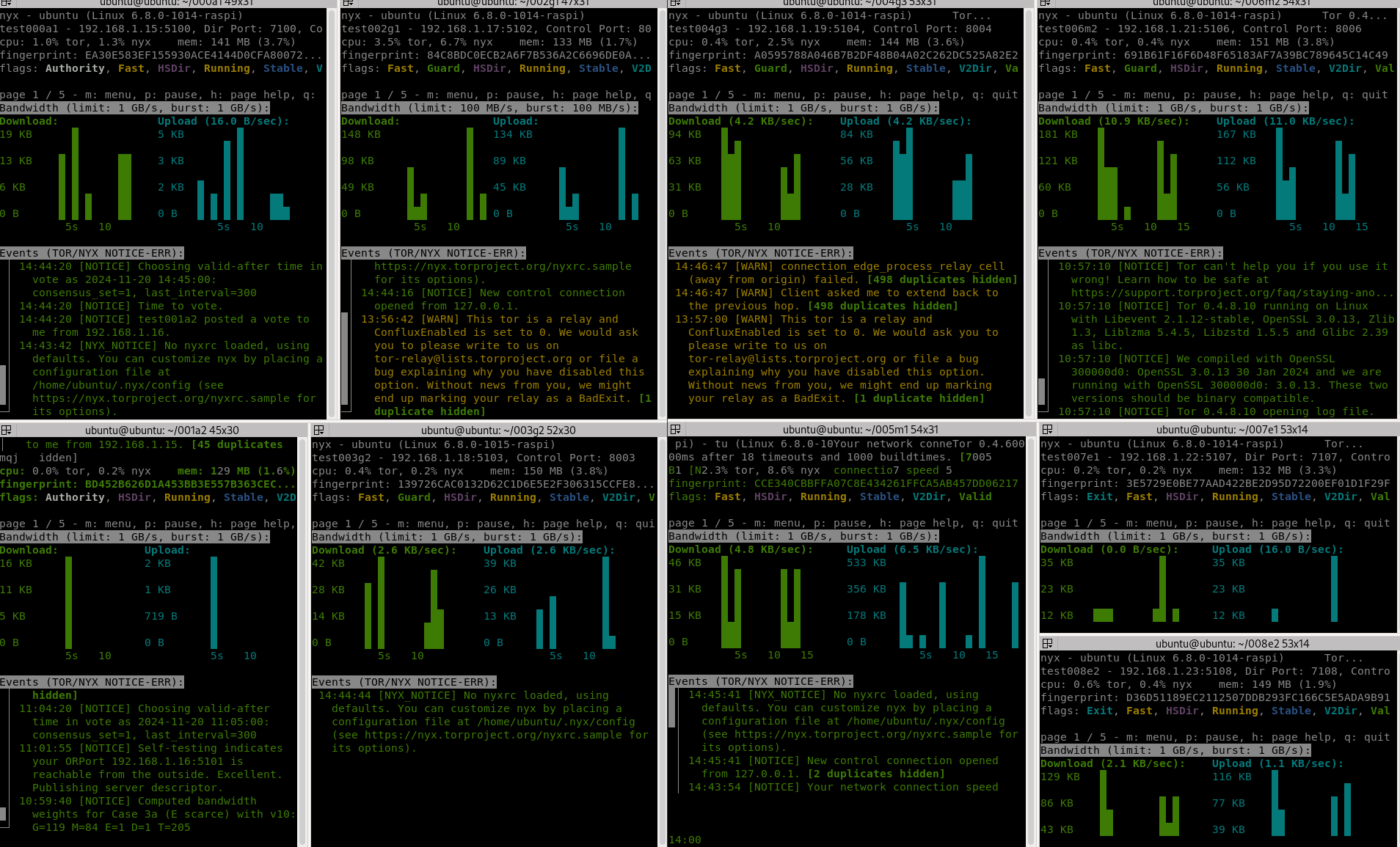}
    \caption{Nyx overview}
    \label{fig:nyxoverview}
\end{figure}

\FloatBarrier


\bibliography{bibliography}
\bibliographystyle{plain}

\end{document}